\documentclass[aps,twocolumn,prb,amsmath,amssymb,superscriptaddress]{revtex4-1}

\usepackage{amsmath}
\usepackage{amssymb}
\usepackage[varg]{txfonts}
\usepackage{color}
\usepackage[colorlinks]{hyperref}
\usepackage{xcolor}
\usepackage{graphicx}
\usepackage{bm}
\usepackage{ulem}
\begin{document}
\date{\today}
\title{Current-induced spin-orbit torque on the surface of a transition metal dichalcogenide connected to the two-dimensional ferromagnet CrI$_3$: Effects of twisting and gating}
\author{Leyla Majidi}
\email{l.majidi@scu.ac.ir}
\affiliation{Department of Physics, Faculty of Science, Shahid Chamran University of Ahvaz, Ahvaz, Iran}
\author{Azadeh Faridi}
\affiliation{School of Quantum Physics and Matter, Institute for Research in Fundamental Sciences (IPM), 19395-5531, Tehran, Iran}
\author{Reza Asgari}
\affiliation{School of Quantum Physics and Matter, Institute for Research in Fundamental Sciences (IPM), 19395-5531, Tehran, Iran}
\affiliation{Department of Physics, Zhejiang Normal University, Jinhua, Zhejiang 321004, China}
\date{\today}

\begin{abstract}
Motivated by recent progress in employing two key classes of two-dimensional materials-topological insulators and transition-metal dichalcogenides (TMDCs)-as spin sources for generating spin-orbit torque (SOT), we investigate current-induced spin polarization and the resulting SOT in bilayers composed of a TMDC (WSe$_2$ or MoSe$_2$) and ferromagnetic chromium iodide (CrI$_3$), beyond the linear response regime. Using the steady-state Boltzmann equation, we find that intraband transitions yield a strong field-like torque on the CrI$_3$ layer, while interband transitions give rise to a comparatively weaker damping-like torque in the WSe$_2$/CrI$_3$ system. Remarkably, the damping-like component is enhanced by up to three orders of magnitude in n-doped MoSe$_2$, reaching a strength comparable to the field-like torque, which itself is an order of magnitude larger than that in the WSe$_2$-based bilayer. Both torque components exhibit strong asymmetry between n-type and p-type doping in WSe$_2$ and MoSe$_2$ systems. Furthermore, we demonstrate that the twist angle plays a crucial role: depending on the TMDC and chemical potential, twisting can reverse the sign of the SOT and significantly modulate its magnitude. Finally, we show that a transverse gate electric field enables substantial tunability of the SOT, by nearly one order of magnitude, and induces a sign reversal at a twist angle of $10.16^{\circ}$.
\end{abstract}
\maketitle
\section{introduction}

Magnetization manipulation and switching mediated by spin torques is the central goal in memory and logic spintronic devices. This has been conventionally achieved by transferring spin angular momentum from a ferromagnetic fixed layer to a ferromagnetic free layer separated by a nonmagnetic spacer applying a perpendicular electric field (spin transfer torque)~\cite{slonczewski1996current,berger1996emission,myers1999current,katine2000current,Zare2017,Majidi2018}. In recent years, an alternative mechanism for generating spin torques has been predicted in nonmagnetic/ferromagnetic bilayers which is based on producing a transverse spin current in the nonmagnetic layer with a strong spin-orbit coupling applying an in-plane electric field. The interface-generated spin current arises from spin-orbit filtering and spin-orbit precession mechanisms~\cite{Amin2018}, and leads to an accumulation of spins at the interface producing a spin torque, the so-called spin-orbit torque (SOT), which can in turn manipulate the magnetic moments in the ferromagnetic layer~\cite{gambardella2011current,manchon2019current}. The spin-orbit filtering current exhibits a fixed spin-polarization direction $\bm{S}$ and generates a torque given by $\bm{m}\times(\bm{m}\times\bm{S})$ on the local magnetization vector $\bm{m}$. In contrast, the spin-orbit precession current possesses a magnetization-dependent spin polarization, expressed as $\bm{m}\times\bm{S}$, and produces a torque of the form $\bm{m}\times(\bm{m}\times[\bm{m}\times\bm{S}])=\bm{m}\times(-\bm{S})$. Generally, these spin torques are classified as damping-like [$\bm{\tau}_{DL}\propto\bm{m}\times(\bm{m}\times\bm{S})$] and field-like ($\bm{\tau}_{FL}\propto\bm{m}\times\bm{S}$), respectively~\cite{manchon2019current}.

Search for source materials with a strong spin-orbit coupling for generating spin-orbit torques have been the subject of an intense study for almost two decades. Large spin-orbit torques have been predicted and observed in a variety of bilayer systems with heavy metal/ferromagnet (HM/FM) and topological insulator/ferromagnet (TI/FM) as the most conventional heterostructures. Considerable spin-orbit torques resulting from strong spin-orbit interaction at both bulk and the interface of HM/FMs with W, Pt and Ta as the most common metals have been reported~\cite{liu2012spin,liu2012current,pai2014enhancement,skinner2015complementary}. On the other hand, TI/FM bilayers mostly based on Sb and Bi chalcogenides have also shown much larger SOT efficiencies with respect to HM/FM bilayers due to the spin-momentum locking of the surface states of topological insulators~\cite{fan2014magnetization,wang2015topological,kondou2016fermi,han2017room} .

\begin{figure}[]
\begin{center}
\includegraphics[width=3in]{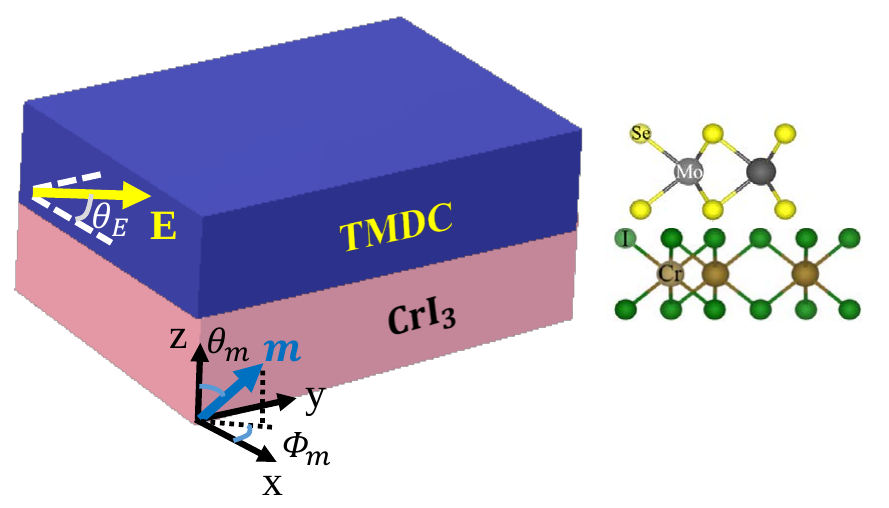}
\end{center}
\caption{\label{Fig:0}(Color online) Schematic of a TMDC/CrI$_3$ bilayer with a magnetization direction $\bm{m}=(\sin\theta_m\cos\phi_m,\sin\theta_m\sin\phi_m,\cos\theta_m)$. There is an applied electric field, $\bm E = |E|(\cos\theta_E,\sin\theta_E,0)$, at the interface of TMDC/CrI$_3$ along $x-y$ plane.}
\end{figure}

Transition metal dichalcogenides (TMDCs) are considered as another promising candidate materials for producing sizable SOT in TMDC/FM bilayers due to their strong intrinsic spin-orbit coupling and diverse symmetry features~\cite{liu2020two,hidding2020spin,tang2021spin,galceran2021control}. The distinct properties of these materials can lead to large charge-to-spin conservation efficiencies and out-of-plane SOTs which are desirable for switching the magnets in bilayers with perpendicular magnetic anisotropy. The large diversity of TMDC materials with different crystalline symmetries and subsequent diverse band structures can lead to diverse spin-orbit torques with conventional and unconventional orientations. In recent years, an increasing number of experimental observations of SOT in several TMDC-based heterostructures have been reported: such as WTe$_2$~\cite{macneill2017control,li2018spin,zhao2020observation} and TaTe$_2$~\cite{stiehl2019current,hoque2020charge}, MoS$_2$~\cite{zhang2016research,shao2016strong} and WSe$_2$~\cite{shao2016strong,novakov2021interface,hidding2021interfacial}, NbSe$_2$~\cite{guimaraes2018spin} and TaS$_2$~\cite{husain2020large} with respectively semi-metallic, semiconducting and metallic features of TMDCs in combination with a variety of ferromagnetic layers. The different and even discrepant results of these experiments show that not only the TMDC materials with different crystalline symmetries and number of layers can lead to diverse spin-orbit torques with conventional and unconventional orientations, but the ferromagnetic layer can also play an active role in this case~\cite{dolui2020spin}.

Besides, spin torque is strongly influenced by the symmetry characteristics of the spin-orbit interaction~\cite{Manchon2009}. In cases where inversion symmetry is preserved- such as impurity-induced spin-orbit interaction or the Luttinger spin-orbit band- the resulting spin torque is a higher-order effect and typically too weak to lead to a reasonable critical switching current density. In contrast, when inversion symmetry is broken, as seen in Rashba and Dresselhaus spin-orbit interactions, the torque emerges as a first-order effect with respect to the SOC parameter and can be effectively utilized to manipulate magnetization direction using relatively low critical switching current densities, on the order of $10^{4}$ to $10^{6}$ A/cm$^2$.

An important issue in applying the SOTs realized in heterostructures for practical purposes is the controllability of the size and the orientation of the spin torques. Although there have been observations of the electric field tunable SOTs in HM/FM~\cite{mishra2019electric}, TI/FM~\cite{han2017room} and TMDC/FM~\cite{lv2018electric,lin2023magnetization} bilayers and also strain-induced unconventional SOTs~\cite{guimaraes2018spin}, the quest for external modulation of SOT is still underway. In this regard, a recent study shows that in a TMDC/FM bilayer, the proximity exchange in the TMDC monolayer (MoSe$_2$ or WSe$_2$) induced by the FM monolayer (CrI$_3$) is widely tunable via gating and twisting~\cite{Zollner19}. This can be a great opportunity to tune the relevant features in this system~\cite{majidi2022electrical,Zollner23}. As CrI$_3$ is one of the most interesting two-dimensional (2D) magnets in which the magnetic moment orientation is ferromagnetic in monolayer and antiferromagnetic in bilayer, a large number of studies have been devoted to its special functionalities~\cite{jiang2018controlling,kim2019exploitable,lei2021magnetoelectric,zhang2022all} as well as its intriguing properties when incorporated into TMDCs~\cite{zhong2017van,seyler2018valley,hu2020manipulation,ge2022enhanced,dolui2020proximity,heibenbuttel2021valley,da̧browski2022all} which stem from the strong proximity effect in these van der Waals (vdW) heterostructures. Several experimental and theoretical studies suggest a strong valley and spin manipulation in TMDC/CrI$_3$ bilayer (mostly WSe$_2$ as TMDC)~\cite{majidi2022electrical,Zollner23,zhong2017van,seyler2018valley,hu2020manipulation,ge2022enhanced}. All-optical switching of the magnetization in CrI$_3$ thin films~\cite{zhang2022all} and also all-optical control of spin in TMDC/CrI$_3$ heterostructures~\cite{da̧browski2022all} are also reported. On the other hand, a recent study shows that a current-driven proximity-induced SOT emerges in a TaSe$_2$/bilayer CrI$_3$ heterostructure which can switch the direction of the magnetization on the bottom monolayer of CrI$_3$ to become parallel to that of the top monolayer, thereby converting bilayer CrI$_3$ from antiferromagnet to ferromagnet\cite{dolui2020proximity}.

Motivated by these findings, here we develop a microscopic theory for the current-induced SOT in TMDC/CrI$_3$ bilayer (with MoSe$_2$ or WSe$_2$ as TMDCs) beyond linear response theory. Of particular interest in this paper is to comment on the tunability of the size and the orientation of the SOT by changing the gate voltage and the twist angle to open a new path for externally controllable spintronic devices. Employing the single-band steady state Boltzmann equation, we find that the main contribution of the intraband transitions to the spin polarization is in linear response regime. The spin polarization originating from the electron occupation changes within intraband transitions generates a strong field-like SOT and the associated intrinsic interband transitions contribute to a weak damping-like SOT on the magnetization of the ferromagnetic CrI$_3$ layer in WSe$_2$-based bilayer, exhibiting asymmetric behavior with respect to the chemical potential and sign changing in the case of n-doped MoSe$_2$ with smaller chemical potentials. Importantly, replacing WSe$_2$ with MoSe$_2$ leads to a significant amplification of the strength of the damping-like term (three orders of magnitude) such that it can be approximately at the same order of the field-like term. The strength of the field-like term is also enhanced by an order of magnitude in MoSe$_2$ in comparison with that of the WSe$_2$-based bilayer.

Remarkably, we demonstrate that twisting will result in a significant amplification of the SOT as well as a sign change in MoSe$_2$- and p-doped WSe$_2$-based bilayers. Moreover, tuning the gate electric field leads to significant enhancement of the strength of the field-like and damping-like SOTs and their sign changes in presence of twisting. Besides, increasing the Rashba spin-orbit coupling enhances the field-like and damping-like SOTs in WSe$_2$- and p-doped MoSe$_2$-based bilayers, while it reduces the strength of the damping-like torque and eliminates the sign change of the field-like torque in n-doped MoSe$_2$ .

The rest of the paper is organized as follows. Section \ref{Model} is devoted to the theoretical model and fundamental formalisms which will be implemented to investigate the current-induced spin-orbit torque in a twisted TMDC/CrI$_3$ bilayer. We present our numerical results for the non-equilibrium spin polarization and the exerted spin-orbit torque on the magnetization of the ferromagnetic layer in Sec. \ref{results}. Finally, a brief summary of results is given in Sec. \ref{conclusion}.

\section{Model and theory}\label{Model}

The device we consider is a TMDC/CrI$_3$ bilayer with monolayer CrI$_3$ as 2D FM and monolayer WSe$_2$ or MoSe$_2$ as TMDC [see Fig. \ref{Fig:0}]. The magnetic insulator CrI$_3$ is weakly coupled to the TMDC by vdW forces, preserving the characteristic electronic band structure of the TMDC. The magnetization direction of the TMDC is found to align with that of the iodine atoms and oppose that of the chromium atoms, resulting in negative proximity exchange parameters when the net magnetization of CrI$_3$ points along the positive z-direction toward the TMDC. The proximity exchange coupling will split the TMDC conduction-band edge by $2|B_c|$ and the valence-band edge by $2|B_v|$, and combined with the intrinsic (valley Zeeman) SOC of the TMDC lifts the valley degeneracy. The effective low-energy Hamiltonian which describes the surface of a TMDC in the presence of proximity exchange~\cite{Zollner19}, has the form
\begin{equation}\label{H}
\mathcal{H}=\mathcal{H}_0+\mathcal{H}_{\Delta}+\mathcal{H}_{soc}+\mathcal{H}_{ex}+\mathcal{H}_{R},
\end{equation}
with
\begin{eqnarray}
&&\mathcal{H}_0=\hbar v_{\rm F}\hat{s}_0\otimes({\chi}_{\tau} {\hat{\sigma}}_x k_x+{\hat{\sigma}}_y k_y),\\
&&\mathcal{H}_{\Delta}=\frac{\Delta}{2}{\hat{s}}_0\otimes{\hat{\sigma}}_{z},\\
&&\mathcal{H}_{soc}={\chi}_{\tau} {\hat{s}}_z\otimes(\lambda_c{\hat{\sigma}}_{+}+\lambda_v{\hat{\sigma}}_{-}),\\
&&\mathcal{H}_{ex}=(\hat{\bm{s}}\cdot\bm{m})\otimes(B_c{\hat{\sigma}}_{+}+B_v{\hat{\sigma}}_{-}),\\
&&\mathcal{H}_{R}=\lambda_R ({\chi}_{\tau} {\hat{s}}_y\otimes{\hat{\sigma}}_{x}- {\hat{s}}_x\otimes{\hat{\sigma}}_{y}).
\end{eqnarray}

The valley index for the K (K') point is ${\chi}_{\tau}=\pm1$ and $v_{\rm F}$ represents the Fermi velocity. The pseudospin Pauli matrices $\hat{\sigma}_i$ ($i = 0,x,y,z$) act on the conduction- and valence-band subspaces and $\hat{s}_i$ ($i = 0,x,y,z$) refers to real spin. The parameter $\Delta$ denotes the orbital gap of the spectrum. The spin-splittings of the conduction (valence) band due to the intrinsic SOC and the proximity exchange are respectively determined by the parameters $\lambda_{c(v)}$ and $B_{c(v)}$. The local magnetization of the ferromagnetic (F) layer is determined by $\bm{m}$. The Rashba SOC parameter $\lambda_R$ is due to the presence of inversion asymmetry in the heterostructure. For short notation, we introduce ${\hat{\sigma}}_{\pm} = ({\hat{\sigma}}_0 \pm {\hat{\sigma}}_z)/2$.

Recently, wide tunability of the proximity exchange in monolayer MoSe$_2$ or WSe$_2$ owing to a ferromagnetic monolayer CrI$_3$ has been explored with respect to twisting and gating~\cite{Zollner19,Zollner23}. In particular, proximity exchange splitting depends on the twist angle between the TMDC and the CrI$_3$. Not only do the magnitudes of the exchange differ, but also the direction of the exchange field for the valence band alters sign. When twisting from 0$^{\circ}$ to 30$^{\circ}$, the  conduction-band proximity exchange parameter is barely influenced, being fixed at around $-1.2$ meV. In contrast, the valence-band proximity exchange parameter is negative for $0^{\circ}$, vanishes and reverses sign at about $8^{\circ}$ ($16^{\circ}$) for WSe$_2$ (MoSe$_2$), and is positive for
a twist angle of $30^{\circ}$. The origin of this reversal is traced back to the twist-angle dependent backfolding of the TMDC K/K' valleys into the CrI$_3$ Brillouin zone and sensitively depends on the atomic stacking configurations of the strained supercells. Moreover, the proximity exchange parameters increase when the applied transverse electric field is turned from negative to positive values, which enables the gate control of the proximity exchange. We should emphasize that the introduced low-energy model Hamiltonian [Eq. (\ref{H})], together with the fitted parameters obtained in Ref. ~\onlinecite{Zollner23}, nicely provide an effective description for the TMDC band edges in twisted TMDC/CrI$_3$ bilayer in the presence of an external electric field. It is worth noting that the gap parameter as well as the Fermi velocity are not affected by external electric fields.

Importantly, large spin-orbit coupling in TMDCs can produce strong enough spin-orbit torque to switch the magnetization of the ferromagnetic layer in TMDC/FM bilayers. The SOT exerting on the ferromagnetic layer, characterized by a local magnetization $\bm{m}=(m_x,m_y,m_z)=(\sin\theta_m\cos\phi_m,\sin\theta_m\sin\phi_m,\cos\theta_m)$, has the form
\begin{equation}
\label{torque}
\bm{\tau}=\frac{2 J}{\hbar}\bm{m}\times\bm{S},
\end{equation}
with $J$ the exchange energy, $\bm{S}=\sum_{\chi}\int\frac{d^2\bm{k}}{(2\pi)^2}\bm{s}^{\chi}(\bm{k})f(\varepsilon_{\bm{k}}^\chi)$ the spin polarization with $\bm{s}^{\chi}(\bm{k}) =({\hbar}/{2})\langle\Psi_{\bm{k}}^\chi|\hat{\bm s}|\Psi_{\bm{k}}^\chi\rangle$ the spin expectation in the $\chi$-th band with eigenstate $\Psi_{\bm{k}}^{\chi}$ and eigenvalue $\varepsilon_{\bm{k}}^{\chi}$, and $f(\varepsilon_{\bm{k}}^{\chi})$ the Fermi-Dirac distribution function. The $\chi$-energy band is defined by $\chi=({\chi}_{\sigma},{\chi}_{s},{\chi}_{\tau})$ in which ${\chi}_{\sigma}=\pm 1$ refers to the conduction and valence band subspaces, and $\chi_s=\pm 1$ labels the two subbands within the conduction and valence bands, which coincide with the spin-subbands in the limit $\lambda_R = 0$.

Applying an in-plane electric field, parallel to the TMDC/FM interface, leads to the spin polarization $\bm{S} = \bm{S}^{oc}+\bm{S}^{in}$ with $\bm{S}^{oc}$ originating from the change of the electron occupation $\delta f(\varepsilon_{\bm{k}}^{\chi})=f(\varepsilon_{\bm{k}}^{\chi})-f^{(0)}(\varepsilon_{\bm{k}}^{\chi})$ within intraband due to an acceleration by the electric field,
\begin{equation}
\label{S_OC}
\bm{S}^{oc}=\sum_{\chi}\int\frac{d^2\bm{k}}{(2\pi)^2}\bm{s}^{\chi}(\bm{k})\delta f(\varepsilon_{\bm{k}}^{\chi}),
\end{equation}
and
\begin{equation}
\label{S_in}
\bm{S}^{in}=\sum_{\chi}\int\frac{d^2\bm{k}}{(2\pi)^2}\delta \bm{s}^{\chi}(\bm{k})f(\varepsilon_{\bm{k}}^{\chi}),
\end{equation}
stemming from the modification of the quasiparticle wave functions~\cite{Kurebayashi,Xiao17,Garate}, where $\delta\bm{s}^{\chi}(\bm{k}) =({\hbar}/{2})Re\langle\Psi_{\bm{k}}^\chi|\bm{\hat{\bm s}}|\delta\Psi_{\bm{k}}^\chi\rangle$ can be traced to the interband contributions in analogy to the intrinsic contribution to the anomalous Hall effect. Therefore, intraband and interband transitions contribute differently to spin polarization. Intraband transitions mainly generate spin currents through scattering processes within the same band linked to field-like torques, while interband transitions involve mixing between different energy bands and are crucial for producing band-structure-driven spin densities tied to damping-like torques.

Employing the single-band steady state Boltzmann equation
\begin{equation}
-\frac{e}{\hbar}\bm{E}\cdot\nabla_{\bm{k}}f(\varepsilon_{\bm{k}}^{\chi})=-\frac{f(\varepsilon_{\bm{k}}^{\chi})-f^{(0)}(\varepsilon_{\bm{k}}^{\chi})}{\gamma(\bm{k})},
\end{equation}
the n-th order non-equilibrium distribution function in the relaxation time approximation $\gamma(\bm{k})=\gamma$, can be obtained from the recursive relations
\begin{equation}
f^{(n)}(\varepsilon_{\bm{k}}^{\chi})=\frac{e\gamma}{\hbar}\bm{E}\cdot\frac{\partial f^{(n-1)}(\varepsilon_{\bm{k}}^{\chi})}{\partial{\bm{k}}}.
\end{equation}
Using this formula, the non-equilibrium distribution function up to the second order can be rewritten as
\begin{eqnarray}
&&f(\varepsilon_{\bm{k}}^{\chi})=f^{(0)}(\varepsilon_{\bm{k}}^{\chi})+f^{(1)}(\varepsilon_{\bm{k}}^{\chi})+f^{(2)}(\varepsilon_{\bm{k}}^{\chi}),\\
&&f^{(0)}(\varepsilon_{\bm{k}}^{\chi})=[e^{\beta(\varepsilon_{\bm{k}}^{\chi}-\mu)}+1]^{-1},\nonumber\\
&&f^{(1)}(\varepsilon_{\bm{k}}^{\chi})=e\gamma\bm{E}\cdot\mathbf{v}^{\chi}\frac{\partial f^{(0)}(\varepsilon_{\bm{k}}^{\chi})}{\partial\varepsilon_{\bm{k}}^{\chi}},\nonumber\\
&&f^{(2)}(\varepsilon_{\bm{k}}^{\chi})=\frac{e^2\gamma^2}{\hbar}[\bm{E}\cdot\frac{\partial(\bm{E}\cdot\mathbf{v}^{\chi})}{\partial\bm{k}}\ \frac{\partial f^{(0)}(\varepsilon_{\bm{k}}^{\chi})}{\partial\varepsilon_{\bm{k}}^{\chi}}+\hbar(\bm{E}.\mathbf{v}^{\chi})^2\frac{\partial^2 f^{(0)}(\varepsilon_{\bm{k}}^{\chi})}{\partial(\varepsilon_{\bm{k}}^{\chi})^2}],\nonumber
\end{eqnarray}
with  $\mathbf{v}^{\chi}=\hbar^{-1}\partial_{\bm{k}}\varepsilon_{\bm{k}}^{\chi}$ the group velocity of charge carriers, $\bm E = (E_x,E_y,0)$ the applied in-plane electric field, $\beta=(k_B T)^{-1}$, $k_B$ the Boltzman constant, $T$ the temperature and $\mu$ the chemical potential.
\begin{table*}[]
\begin{center}
\begin{tabular}{l|c|c|c|c|c|r}
\hline\hline
\vline ~ & & & & & & \vline\\
\vline ~ \textsl{Structure} & \textsl{$\Delta$[eV]}& \textsl{$v_{\rm F}$[$10^5$ m/s]} & \textsl{$\lambda_c$[meV]} & \textsl{$\lambda_v$[meV]} & \textsl{$B_c$[meV]} & \textsl{$B_v$[meV]} \vline \\
\vline ~ & & & & & & \vline\\
\hline
\vline ~ & & & & & & \vline\\
\vline ~ WSe$_2$/CrI$_3$ & 1.403&5.873&13.81&240.99&-1.460&-0.918 \vline\\
\vline ~ (no twist)& & & & & & \vline\\
\hline
\vline ~ & & & & & & \vline\\
\vline ~ WSe$_2$/CrI$_3$ & 1.477  &5.940& 13.81&240.99 & -1.226&1.525 \vline\\
\vline ~ (with twist $30^\circ$)& & & & & & \vline\\
\hline
\vline ~ & & & & & & \vline\\
\vline ~ MoSe$_2$/CrI$_3$ & 1.343&4.358& -9.678 & 94.43 & -1.343& -0.901 \vline\\
\vline ~ (no twist) & & & & & & \vline\\
\vline ~ & & & & & & \vline\\
\hline
\vline ~ & & & & & & \vline\\
\vline ~ MoSe$_2$/CrI$_3$ & 1.401&4.622& -9.678& 94.43 & -1.176& 0.484 \vline\\
\vline ~ (with twist $30^\circ$)& & & & & & \vline\\
\hline\hline
\end{tabular}
\caption{The orbital gap $\Delta$, Fermi velocity $v_{\rm F}$, spin-orbit coupling $\lambda_{c(v)}$ and the proximity exchange $B_{c(v)}$ of the conduction (valence) band for TMDC/CrI$_3$ bilayer (with WSe$_2$ and MoSe$_2$ as TMDC) in the absence or presence of the twist angle 30$^{\circ}$ [see Ref. ~\onlinecite{Zollner23}]. Notice that $B_{v}$ changes sign in twisted cases.}
\label{table1}
\end{center}
\end{table*}
First, we calculate the spin polarization stemming from intraband transitions, using Eq. (\ref{S_OC}). Diagonalizing the effective Hamiltonian [Eq. (\ref{H})], the eigenstates at a given energy $\varepsilon_{\bm{k}}^{\chi}$ can be obtained as
\begin{equation}
\label{psi}
\Psi_{\bm{k}}^{\chi}=A_{\chi}^{{\chi}_{\tau}}\ e^{i {\chi}_{\tau} {\chi}_{\sigma} k_x x} e^{i {\chi}_{\sigma} k_y y}
\left(
\begin{array}{c}
{\chi}_{\tau} {\chi}_{\sigma} N_{\chi}^{{\chi}_{\tau}}\ e^{-i{\chi}_{\tau}\theta}\\
1\\
i P_{\chi}^{{\chi}_{\tau}}\\
i {\chi}_{\tau} {\chi}_{\sigma} M_{\chi}^{{\chi}_{\tau}}\ e^{i{\chi}_{\tau}\theta}
\end{array}
\right)
\end{equation}
for the $\chi$-band with momentum-energy relation
\begin{eqnarray}
\label{dispersion}
|\bm{k}_{\chi}^{{\chi}_{\tau}}|&=&\frac{1}{\sqrt{2}\hbar v_{\rm F}} [a_1 a_2 + a_3 a_4 - (a_1 + a_2 + a_3 + a_4)\ \varepsilon_{\bm{k}}^{\chi} + 2\ {\varepsilon_{\bm{k}}^{\chi}}^2\nonumber\\
&+& \chi_s\ [(a_1 a_2- a_3 a_4 + (-a_1 - a_2 + a_3 + a_4)\ \varepsilon_{\bm{k}}^{\chi})^2 \nonumber\\
&+& 16\ (a_{1(2)}- \varepsilon_{\bm{k}}^{\chi}) (a_{4(3)} - \varepsilon_{\bm{k}}^{\chi}) \lambda_R^2]^{1/2}]^{1/2}.
\end{eqnarray}
Here,
\begin{eqnarray}
&&N_{\chi}^{+}=\frac{\hbar v_{\rm F} |\bm{k}_{\chi}^{+}|}{\varepsilon_{\bm{k}}^{\chi}-a_1},\nonumber\\
&&P_{\chi}^{+}=\frac{\varepsilon_{\bm{k}}^{\chi}-a_2-\hbar v_{\rm F} |\bm{k}_{\chi}^{+}| N_{\chi}^{+}}{2\lambda_R},\nonumber\\
&&M_{\chi}^{+}=\frac{\hbar v_{\rm F} |\bm{k}_{\chi}^{+}| P_{\chi}^{+}}{\varepsilon_{\bm{k}}^{\chi}-a_4},\nonumber
\end{eqnarray}
and
\begin{eqnarray}
&&N_{\chi}^{-}=\frac{\varepsilon_{\bm{k}}^{\chi}-a_2}{\hbar v_{\rm F} |\bm{k}_{\chi}^{-}|},\nonumber\\
&&M_{\chi}^{-}=\frac{N_{\chi}^{-} (a_1-\varepsilon_{\bm{k}}^{\chi})+\hbar v_{\rm F} |\bm{k}_{\chi}^{-}|}{2\lambda_R}\ e^{-2 i{\chi}_{\tau}\theta},\nonumber\\
&&P_{\chi}^{-}=\frac{\hbar v_{\rm F} |\bm{k}_{\chi}^{-}| M_{\chi}^{-}}{\varepsilon_{\bm{k}}^{\chi}-a_3},\nonumber
\end{eqnarray}
with superscript $\pm$ denoting the K (K') valley with ${\chi}_{\tau}=\pm 1$, $a_{1(3)}=\pm{\chi}_{\tau}\lambda_{c}\pm B_{c} \cos\theta_m+\Delta/2$, $a_{2(4)}=\pm{\chi}_{\tau}\lambda_{v}\pm B_{v} \cos\theta_m-\Delta/2$, $\theta=\arctan(k_y/k_x)$, and  $A_{\chi}^{{\chi}_{\tau}}=({|N_{\chi}^{{\chi}_{\tau}}|}^2+{|P_{\chi}^{{\chi}_{\tau}}|}^2+{|M_{\chi}^{{\chi}_{\tau}}|}^2+1)^{-1/2}$. Note that the local magnetization $\bm{m}$ is assumed to lie along the z-direction, with azimuthal and polar angles $\phi_m = 0$ and $\theta_m = 0$, respectively.

Using Eqs. (\ref{psi}) and (\ref{dispersion}), the spin expectation components are obtained as follows,
\begin{equation}
\label{spinexpectation1}
s_x^{\chi}=-{\hbar}{\chi}_{\tau}\ {\chi}_{\sigma}\ |A_{\chi}^{{\chi}_{\tau}}|^2[\rm{Im}({N_{\chi}^{{\chi}_{\tau}}}^*\ P_{\chi}^{{\chi}_{\tau}}\ e^{i{\chi}_{\tau}\theta})+\rm{Im}({M_{\chi}^{{\chi}_{\tau}}}e^{i{\chi}_{\tau}\theta})],
\end{equation}
\begin{equation}
\label{spinexpectation2}
s_y^{\chi}={\hbar}{\chi}_{\tau}\ {\chi}_{\sigma}\ |A_{\chi}^{{\chi}_{\tau}}|^2\ [\rm{Re}({N_{\chi}^{{\chi}_{\tau}}}^*\ P_{\chi}^{{\chi}_{\tau}}\ e^{i{\chi}_{\tau}\theta})+\rm{Re}({M_{\chi}^{{\chi}_{\tau}}}\ e^{i{\chi}_{\tau}\theta})],
\end{equation}
\begin{equation}
\label{spinexpectation3}
\hspace{-21mm}s_z^{\chi}={\hbar}\ |A_{\chi}^{{\chi}_{\tau}}|^2\ (|N_{\chi}^{{\chi}_{\tau}}|^2-|P_{\chi}^{{\chi}_{\tau}}|^2-|M_{\chi}^{{\chi}_{\tau}}|^2+1)/{2}.
\end{equation}

Also, the non-equilibrium distribution functions can be written in terms of the excitation energy, $\varepsilon=\varepsilon_{\bm{k}}^{\chi}-\mu$, as follows
\begin{eqnarray}
f^{(0)}(\varepsilon)&=&[e^{\beta \varepsilon}+1]^{-1},\\
f^{(1)}(\theta,\varepsilon)&=&-\beta \gamma e|\bm E| (\mathrm{v}_x^{\chi}\cos\theta_E +\mathrm{v}_y^{\chi}\sin\theta_E) e^{\beta \varepsilon}[f^{(0)}(\varepsilon)]^2,\\
f^{(2)}(\theta,\varepsilon)&=&(\beta \gamma e|\bm E|)^2([(\partial \mathrm{v}_x^{\chi}/\partial k_x)\ {\cos^2\theta_E}+ (\partial \mathrm{v}_y^{\chi}/\partial k_y)\ \sin^2\theta_E\nonumber\\
&+&(\partial \mathrm{v}_x^{\chi}/\partial k_y+\partial \mathrm{v}_y^{\chi}/\partial k_x)\cos\theta_E \sin\theta_E](\partial f^{(0)}(\varepsilon)/\partial\varepsilon)\nonumber\\
&+&[\mathrm{v}_x^{\chi} \cos\theta_E +\mathrm{v}_y^{\chi}\sin\theta_E]^2(\partial^2 f^{(0)}(\varepsilon)/\partial\varepsilon^2)),
\end{eqnarray}
with $\bm{\mathrm{v}}^{\chi}={\chi}_{\sigma}{[\hbar\ |\bm{k}_{\chi}(\varepsilon)|\partial |\bm{k}_{\chi}(\varepsilon)|/\partial \varepsilon]}^{-1}(k_x,k_y)={\chi}_{\sigma}{[\hbar\ \partial |\bm{k}_{\chi}(\varepsilon)|/\partial \varepsilon]}^{-1}(\cos\theta,\sin\theta)$, ${\chi}_{\sigma}=sgn(\mu+\varepsilon)$, $\theta_E=\arctan(E_y/E_x)$, and $|\bm E|=\sqrt{E_x^2+E_y^2}$.

Finally, the intraband spin polarization $\bm{S}^{oc}$ in response to an in-plane applied electric field, and the resulting spin-orbit torque $\bm{\tau}^{oc}=2 J{\hbar}^{-1}\bm{m}\times\bm{S}^{oc}$, exerting on the ferromagnetic layer characterized by a local magnetization $\bm{m}=\cos\theta_m\ \hat{z}$ and $J$ the exchange energy, can be obtained using
\begin{widetext}
\begin{equation}\label{S_oc_2}
\bm{S}^{oc}=(\frac{1}{2\pi})^2\sum_{\chi}\int_0^\infty d\varepsilon\ |\bm{k}_{\chi}^{\tau}(\varepsilon)| \frac{\partial |\bm{k}_{\chi}^{\tau}(\varepsilon)|}{\partial \varepsilon}\bm{s}^{\chi}(\varepsilon)\int_{0}^{2\pi} d\theta\ \ [f^{(1)}(\theta,\varepsilon)+f^{(2)}(\theta,\varepsilon)].
\end{equation}
\end{widetext}
The contribution of the linear $\bm{S}^{oc(1)}=\sum_{\chi}\int\frac{d^2\bm{k}}{(2\pi)^2}\bm{s}^{\chi}(\bm{k}) f^{(1)}(\varepsilon_{\bm{k}}^{\chi})$ and non-linear $\bm{S}^{oc(2)}=\sum_{\chi}\int\frac{d^2\bm{k}}{(2\pi)^2}\bm{s}^{\chi}(\bm{k}) f^{(2)}(\varepsilon_{\bm{k}}^{\chi})$ terms to the intraband spin polarization $\bm{S}^{oc}$ are presented in Appendix \ref{sec:appendix A} [see Fig. \ref{Fig:6}]. We find that the contribution of the non-linear term $\bm{S}^{oc(2)}$ is $10^{-4}$ times smaller than that of the linear term. Therefore, the spin polarization $\bm{S}^{oc}$ and the resulting spin-orbit torque mainly arise from the linear term.

Then, we evaluate the contribution of the intrinsic interband transitions on the spin polarization in the linear response regime, by making use of Eq. (\ref{S_in}). Following the perturbation method to find the modifications in the wave function, we have~\cite{Kurebayashi,Xiao17,Garate}
\begin{equation}\label{S_in_2}
{\bm S}^{in}=\frac{e\hbar^2}{A} \sum_{\chi\neq\chi',\bf{k}}[f^{(0)}(\varepsilon^{\chi}_{\bf k})-f^{(0)}(\varepsilon^{\chi'}_{\bf k})]\frac{\rm{Im}\bigl[\langle \Psi^{\chi}_{\bf k}\vert{\hat{\bm s}}\vert\Psi^{\chi'}_{\bf k}\rangle\langle \Psi^{\chi'}_{\bf k}\vert{\bf{\rm{\hat{v}}\cdot E}}\vert\Psi^{\chi}_{\bf k}\rangle\bigr]}{(\varepsilon^{\chi}_{\bf k}-\varepsilon^{\chi'}_{\bf k})^2}
\end{equation}
for each valley with $\rm{\hat{v}}=\hbar^{-1}\partial_{\bm{k}}\mathcal{H}$. Finding the imaginary part on the right hand side (rhs) of the above equation and performing the integration over $\theta$, we find $S^{in}_{z,{\chi}_{\tau}}=0$ and the $x$ and $y$ contributions of the intrinsic term of the spin polarization are given by
\begin{eqnarray}\label{Sx}
&S^{in}_{x(y),\ {\chi}_{\tau}}&=(2\pi e v_F \hbar^2)\ {\chi}_{\tau}\ \!\!\!\sum_{\chi\neq\chi'} \nonumber\\
&\times&\int\!\! \!  kdk \frac{[f^{(0)}(\varepsilon^{\chi}_{\bf k})\!-\!f^{(0)}(\varepsilon^{\chi'}_{\bf k})]}{(\varepsilon^{\chi}_{\bf k}-\varepsilon^{\chi'}_{\bf k})^2}
(A_{\chi}^{{\chi}_{\tau}}A_{\chi'}^{{\chi}_{\tau}}\!)^2(\!\alpha^{{\chi}_{\tau}}\!\beta'^{{\chi}_{\tau}}\!\!-\!\beta^{{\chi}_{\tau}}\!\alpha'^{{\chi}_{\tau}}\!)E_{x(y)},\nonumber\\
\end{eqnarray}
where $\alpha^{{\chi}_{\tau}}=(N_{\chi}^{{\chi}_{\tau}}P_{\chi'}^{{\chi}_{\tau}}+M_{\chi'}^{{\chi}_{\tau}}$), $\beta^{{\chi}_{\tau}}=(P_{\chi}^{{\chi}_{\tau}}N_{\chi'}^{{\chi}_{\tau}}+M_{\chi}^{{\chi}_{\tau}})$, $\alpha'^{{\chi}_{\tau}}=(M_{\chi}^{{\chi}_{\tau}}P_{\chi'}^{{\chi}_{\tau}}+N_{\chi'}^{{\chi}_{\tau}})$, and $\beta'^{{\chi}_{\tau}}=(P_{\chi}^{{\chi}_{\tau}}M_{\chi'}^{{\chi}_{\tau}}+N_{\chi}^{{\chi}_{\tau}})$.
\section{Numerical results and discussion}\label{results}

In this section, we present our numerical results for the current-induced spin-orbit torque exerted on the magnetization of the ferromagnetic layer, obtained by using Eqs. (\ref{S_oc_2}) and (\ref{S_in_2}). Before evaluating our numerical results, it is worth noting that the orbital gap $\Delta$, spin-orbit coupling $\lambda_{c(v)}$, and the exchange interaction $B_{c(v)}$ parameters for different TMDCs (like WSe$_2$ and MoSe$_2$), in the absence or presence of twisting are set to the values obtained in Ref. ~\onlinecite{Zollner23} [see Table. \ref{table1} for the twist angles 0$^\circ$ and 30$^\circ$]. It has been shown that the absence or presence of the SOC term in the bilayer calculations does not lead to significant changes in the magnitude of the exchange parameters. However, twisting the CrI$_3$ layer relative to the WSe$_2$ (MoSe$_2$) changes the sign of the exchange field in the valence band around 8$^\circ$ (16$^\circ$) and makes the valence-band spin-splitting opposite in sign, in the absence of SOC. Therefore, twisting can remain an effective tool to modify the proximity exchange field.

The chemical potential $\mu$ is in units of electron volt (eV). For practical applications in electronic devices, the single-layer and multi-layer TMDCs can be $n$- or $p$-doped on generating desirable charge carriers~\cite{Radisavljevic,Fontana,Laskar}. We set the relaxation time $\gamma = 3$ ps, the magnitude of the in-plane electric field $E=0.2$ mV/nm, $k_B T=25.7$ meV for room temperature, and replace the exchange parameter $J$ with $B_{c(v)}$ for the conduction (valence) band.

\begin{figure*}[t]
\begin{center}
\includegraphics[width=5.5in,height=5.5in]{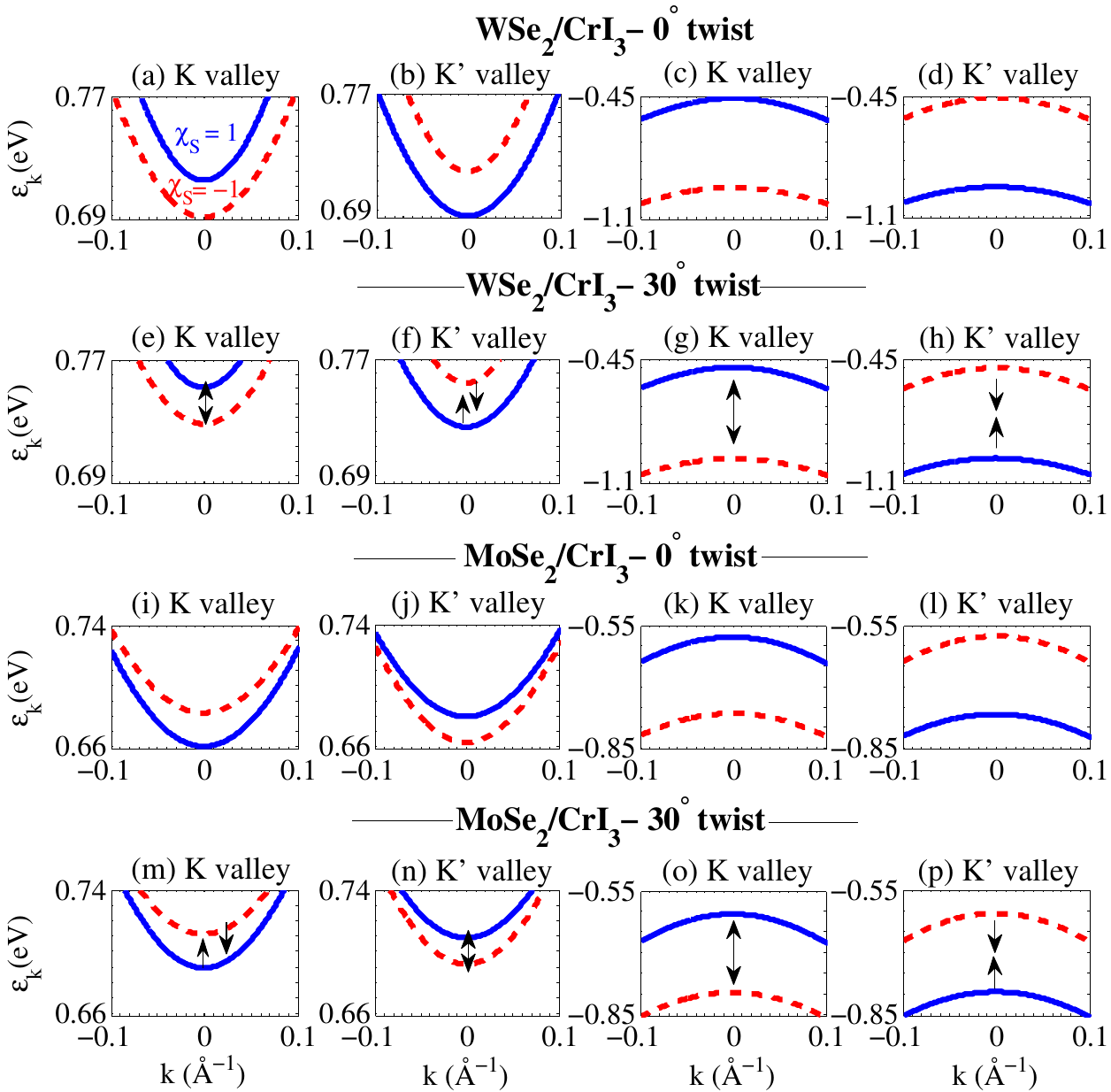}
\end{center}
\caption{\label{Fig:00}(Color online) Panels (a)-(h) and (i)-(p) present the zoomed-in conduction- and valence-band edges of the low-energy band structure for WSe$_2$/CrI$_3$ and MoSe$_2$/CrI$_3$ bilayers, respectively, around the K and K' valleys (${\chi}_{\tau}=\pm1$) at twist angles $0^{\circ}$ and $30^{\circ}$. The two subbands with $\chi_s=1$ and $-1$ in the conduction and valence bands are respectively indicated by blue solid and red dashed lines, following the labeling convention introduced in panel (a). Arrows mark the relative displacement of electronic subbands in the twisted bilayer compared to the untwisted case. The magnetization direction is set to $\theta_m=0$, and Rashba SOC parameter is $\lambda_R=1$ meV.}
\end{figure*}

First, we evaluate the effect of twisting on the low-energy band structure of the TMDC/CrI$_3$ bilayer. In pristine TMDC monolayers, the K and K' valleys are degenerate due to time-reversal symmetry. Proximity to the two-dimensional magnetic material CrI$_3$ offers a direct and effective mechanism to lift this valley degeneracy via the magnetic proximity effect. In the case of WSe$_2$/CrI$_3$ heterostructure, when there is no Rashba SOC, the presence of a negative proximity-induced exchange field shifts the spin-up subbands downward and the spin-down subbands upward in the conduction and valence bands, respectively, by $|B_c|$ and $|B_v|$. Therefore, the spin-subbands of the K valley are brought close to each other and those of the K' valley are got farther from each other. In contrast, the spin-subbands in the band structure of the MoSe$_2$/CrI$_3$ bilayer, are different from those of the WSe$_2$/CrI$_3$ heterostructure because of the different sign of $\lambda_c$. The presence of the proximity exchange field brings the spin-up subbands of each valley as well as spin-down subbands closer together. Activating the Rashba SOC mixes the spin-up and spin-down subbands. The resulting subbands labeled by $\chi_s = \pm 1$, which reduce to spin-polarized states in the limit $\lambda_R = 0$, undergo only slight energy shifts owing to the small Rashba parameter. The corresponding low-energy band structures of WSe$_2$- and MoSe$_2$-based bilayers are presented respectively in Figs. \ref{Fig:00}(a)-\ref{Fig:00}(d) and \ref{Fig:00}(i)-\ref{Fig:00}(l).

\begin{figure}[t]
\begin{center}
\includegraphics[width=3.4in]{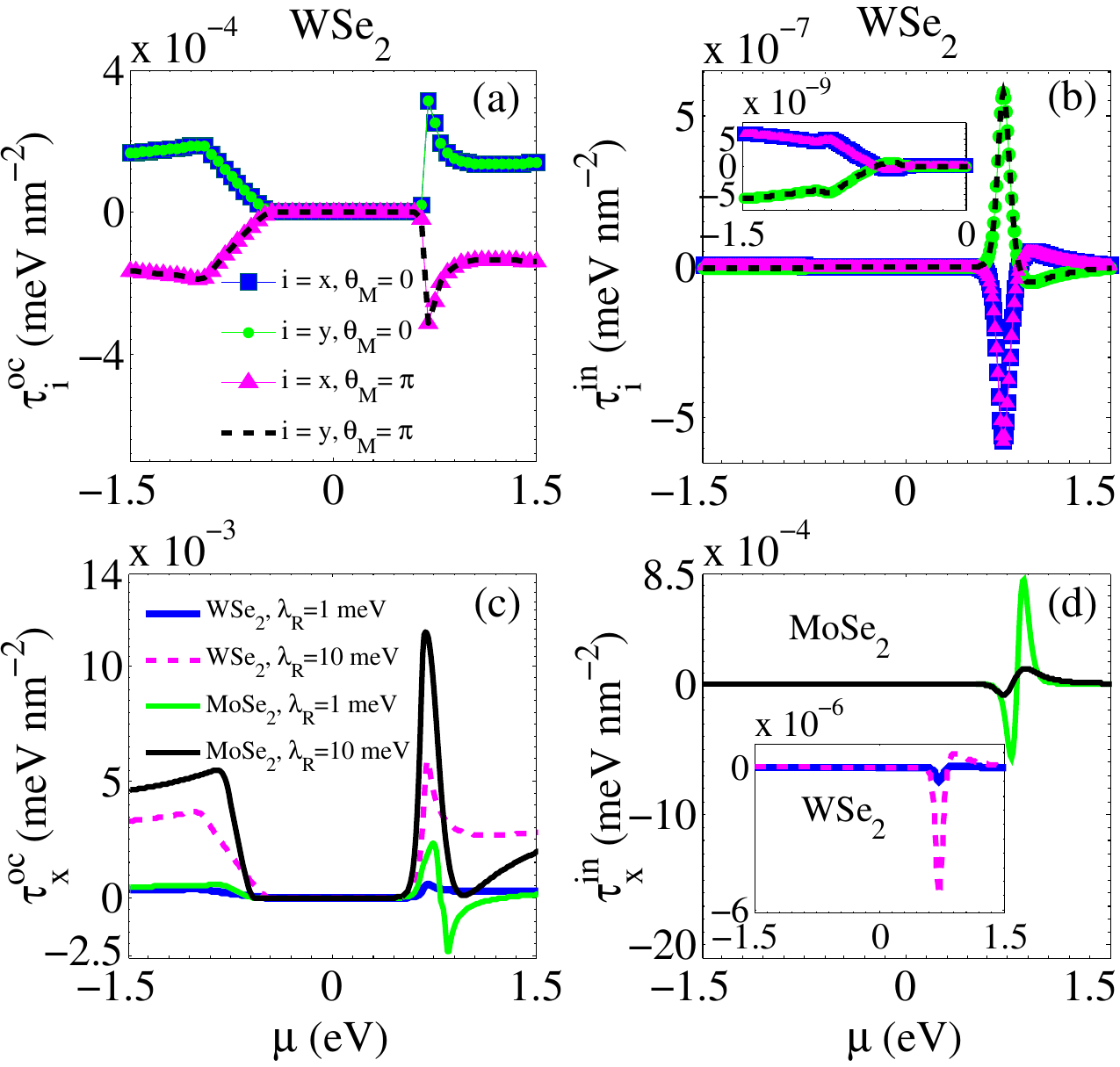}
\end{center}
\caption{\label{Fig:2} (Color online) Chemical potential dependence of (a) intraband $\bm{\tau}^{oc}$ and (b) intrinsic interband $\bm {\tau}^{in}$ spin-orbit torque components in WSe$_2$/CrI$_3$ bilayer for two magnetization directions $\theta_m=0$ and $\pi$, when $\lambda_R = 1$ meV, $\theta_{twist}=0^\circ$ and $\theta_E=\pi/4$. Panels (c) and (d) respectively show ${\tau}_x^{oc}$ and ${\tau}_x^{in}$ as functions of the chemical potential $\mu$ for two values of Rashba parameter in the untwisted WSe$_2$- and MoSe$_2$-based bilayer, when $\theta_m=0$. Inset of (b) shows the zoomed-in view of ${\tau}_i^{in}$ in the case of p-doped bilayer. The labels in panels (b) and (d) follow the same conventions as those in panels (a) and (c), respectively. Note that we have ${\tau}_y^{oc}={\tau}_x^{oc}$ in the case of $\theta_E=\pi/4$.}
\end{figure}

Twisting the CrI$_3$ layer relative to the TMDC monolayer from $0^\circ$ to $30^\circ$ induces a sign reversal in the proximity-induced exchange field within the valence band, occurring near $8^\circ$ for WSe$_2$ and $16^\circ$ for MoSe$_2$. In the absence of SOC, this sign reversal flips the valence-band spin-splitting. Therefore, the TMDC conduction-band edge remains nearly unchanged at around $-3$ meV and the valence-band edge spin-splitting varies almost linearly with twist angle, ranging from $-2$ to $2$ meV~\cite{Zollner23}. In the presence of SOC, however, the twist-angle-induced reversal of the exchange field in the valence band does not modify the spin character of the subbands, since the exchange-driven splittings in both conduction and valence bands are at least three orders of magnitude smaller than the intrinsic SOC, rendering the exchange effect insufficient to alter the spin index. Accordingly, the two $\chi_s = \pm 1$ subbands in the conduction or valence bands of the WSe$_2$/CrI$_3$ heterostructure shift away from each other at the K valley-except at the twist angle $\theta_{twist} = 10.16^{\circ}$ in the conduction band- while they move closer together at the K' valley, with the same exception at $\theta_{twist} = 10.16^{\circ}$ [see Figs. \ref{Fig:00}(e)-\ref{Fig:00}(h) for the twist angle $\theta_{twist}=30^{\circ}$]. In contrast, in the MoSe$_2$/CrI$_3$ heterostructure, the conduction-band subbands shift closer together at the K valley and farther apart at the K' valley, with a similar anomaly at $\theta_{twist} = 10.16^{\circ}$, while those of the valence band show a reversed trend [see Figs. \ref{Fig:00}(m)-\ref{Fig:00}(p) for $\theta_{twist}=30^{\circ}$]. The twist-angle dependence of the $\chi_s = \pm 1$ subband edge energies in the conduction and valence bands of the K valley, together with the corresponding splitting energies, are presented in Appendix \ref{sec:appendix B} for WSe$_2$- and MoSe$_2$-based bilayers. The splitting energies of the $\chi_s = 1$ and $-1$ subbands in the conduction and valence bands of WSe$_2$, as well as in the valence (conduction) band of MoSe$_2$, increase (decrease) almost linearly with the twist angle, consistent with the behavior of the proximity exchange coupling in the conduction and valence bands reported in Ref. \onlinecite{Zollner23} [see Fig. \ref{Fig:9}]. The anomalous behavior at $\theta_{twist} = 10.16^{\circ}$ is attributed to the enhancement of the strength of the proximity exchange field in the conduction band. Besides, both the Fermi velocity $v_{\rm F}$ and the orbital gap $\Delta$ are affected by twisting; their values vary with the twist angle and may either increase or decrease accordingly.

Now, we are in a position to investigate the contributions of intraband and interband transitions to the current-induced spin polarization and the resulting SOT acting on the ferromagnetic layer. As discussed previously, intraband processes yield extrinsic, scattering-driven terms, while interband transitions provide intrinsic, band-structure-driven contributions. Hereupon, intraband effects are governed by the Fermi velocity and spin expectation values of carriers, whereas interband transitions depend on band topology and are linked to band gaps, spin-splitting, and symmetry breaking. Crucially, the SOT components depend not only on band eigenvalues but also on the distinct geometrical features encoded in the eigenstates. Further to our earlier discussion on tunability of the proximity-induced exchange field in TMDC/CrI$_3$ bilayers, in what follows, we analyze the intra and interband contributions to SOT focusing on three key tuning factors: doping, twist angle, and gate electric field.

$Doping:$Figures \ref{Fig:2}(a) and \ref{Fig:2}(b) present the applied current-induced in-plane spin-orbit torque $\bm{\tau}^{oc}$ and $\bm{\tau}^{in}$ components, originating from intraband and interband transitions, respectively, as functions of the chemical potential $\mu$ in the untwisted WSe$_2$/CrI$_3$ heterostructure for $\theta_E=\pi/4$ . The components ${\tau}_{x(y)}^{oc}$ are odd with respect to the magnetization direction $\theta_m$, indicating that ${\tau}^{oc}$ behaves as a field-like torque. In contrast, the intrinsic interband contributions are even under magnetization reversal, consistent with a damping-like character.
The $x$-component $\tau_x^{oc}$ (equal to $\tau_y^{oc}$ for $\theta_E = \pi/4$) exhibits a strongly asymmetric dependence on $\mu$, featuring a gap originated from the absence of quasiparticle states within the band gap of the TMDC/CrI$_3$ bilayer and a sharp peak for the n-doped structure, without any sign reversal in the WSe$_2$/CrI$_3$ bilayer. Substituting MoSe$_2$ as the TMDC significantly enhances the ${\tau}^{oc}$ magnitude and induces a sign change in the n-doped MoSe$_2$ structure [see Fig. \ref{Fig:2}(c)]. While the intrinsic spin-orbit torque ${\tau}^{in}$, in both WSe$_2$- and MoSe$_2$-based bilayers, is much weaker in the p-doped regime compared to the n-doped case [see Figs. \ref{Fig:2}(b) and \ref{Fig:2}(d)]. Notably, the damping-like term ${\tau}^{in}$ in MoSe$_2$ exceeds that in WSe$_2$ and is comparable in magnitude to the field-like term ${\tau}^{oc}$. To understand this behavior, it is important to note that the intrinsic spin-orbit torque arises from interband contributions [see Eq.~\eqref{S_in_2}]. Since its magnitude is inversely proportional to the energy separation between bands, the intrinsic torque is generally expected to become larger when the subbands are closer in energy. In these heterostructures, the intrinsic SOC parameter is primarily responsible for the spin splitting. Comparing the intrinsic SOC parameters of MoSe$_2$/CrI$_3$ ($\lambda_c = -9.678$ meV and $\lambda_v = 94.43$ meV) with those of WSe$_2$/CrI$_3$ ($\lambda_c = 13.81$ meV and $\lambda_v = 240.99$ meV), we observe that the SOC strength is substantially larger in the valence bands than in the conduction bands for both systems. Consequently, the valence subbands exhibit much stronger spin-splitting than the conduction subbands, which explains why the intrinsic torque in the p-doped regime is considerably weaker than in the n-doped case [see Figs.~\ref{Fig:00} and \ref{Fig:9}]. A similar trend is observed when comparing the n-doped WSe$_2$-based system with the n-doped MoSe$_2$-based system. In the latter case, the two conduction subbands lie closer in energy, particularly at larger $k$ values, which enhances the interband mixing and consequently leads to a larger intrinsic torque. Variation of the chemical potential further reveals a sign reversal of ${\tau}^{in}$, in n-doped systems which arises from the opposite spin character and Berry curvature structure of the two spin-split subbands. As the chemical potential moves between these subbands under doping, the occupation difference and geometric interband matrix elements change sign, reversing the direction of the interband spin polarization.

In order to explain the behavior of the intraband spin-orbit torque $\bm{\tau}^{oc}$ in more details, we analyze the contribution of the carriers from $\chi_s = \pm 1$ subbands of the K and K' valleys to the net spin polarization [see Appendix \ref{sec:appendix D}]. We find that the spin polarization of the two subbands within each valley are opposite in sign.  The locations of the peaks coincide with the energies where the spin-split subbands approach each other most closely in momentum space [see Figs. \ref{Fig:11}(a) and \ref{Fig:11}(c) respectively for MoSe$_2$- and WSe$_2$-based bilayers]. Thereupon, the chemical potential dependence of the resulting spin polarization in a single valley can manifest in two distinct forms, dictated by the magnitudes and signs of the intrinsic spin-orbit coupling parameters $\lambda_c$ and $\lambda_v$. In MoSe$_2$/CrI$_3$ bilayer, the negative $\lambda_c$ parameter produces a sharp, high-amplitude peak in the spin polarization for each subband. The resultant sum of the opposing spin polarizations from the $\chi_s = \pm 1$ subbands leads to a sign change in the spin polarization of the two valleys and accordingly the net spin polarization of the n-doped systems [see Figs. \ref{Fig:11}(a) and \ref{Fig:11}(b)]. While in WSe$_2$-based structure, the positive $\lambda_c$ results in a broad peak with low-amplitude in each subband and therefore suppresses the sign change in the spin polarization [see Figs. \ref{Fig:11}(c) and \ref{Fig:11}(d)]. Additionally, the larger $\lambda_c$ and, in particular, $\lambda_v$ values in WSe$_2$ relative to MoSe$_2$ lead to a diminished spin polarization, consequently promoting a stronger intraband SOT in MoSe$_2$-based bilayer [see Fig. \ref{Fig:12}(a)]. According to the Hamiltonian [Eq. (\ref{H})], the intrinsic SOC term-which carries a valley index- and the proximity-induced exchange term together generate an effective out of plane field $B_{z,c(v)}^{eff}=\chi_{\tau}\lambda_{c(v)}+B_{c(v)}$ in the conduction (valence) band. In contrast, the Rashba SOC term produces an in-plane effective field $\bm{B}_R(\bm{k})=\lambda_R(-\sigma_y^{\chi}(\bm{k}),{\chi}_{\tau}\ \sigma_x^{\chi}(\bm{k}),0)$, whose components are determined by the pseudospin expectation values ${\sigma}_{x(y)}^{\chi}(\bm{k})=\langle\Psi_{\bm{k}}^\chi|{\hat{\sigma}_{x(y)}}|\Psi_{\bm{k}}^\chi\rangle$.  We have found (not shown) that the in-plane components of the pseudospin expectation value, obtained as ${\sigma}_{x(y)}^{\chi}(\bm{k})=\pm 2\ {\chi}_{\tau}\ {\chi}_{\sigma}\ |A_{\chi}^{{\chi}_{\tau}}|^2\ (\rm{Re}(\rm{Im})[{N_{\chi}^{{\chi}_{\tau}}}\ e^{-i{\chi}_{\tau}\theta}]+\rm{Re}(\rm{Im})[{{M}_{\chi}^{{\chi}_{\tau}}}^*\ P_{\chi}^{{\chi}_{\tau}}\ e^{-i{\chi}_{\tau}\theta}])$, are larger in MoSe$_2$-based bilayer. Their magnitude increases with the chemical potential, and eventually saturates at high doping levels. Because MoSe$_2$ possesses smaller intrinsic SOC parameters than WSe$_2$, the resulting out of plane effective field $B_{z,c(v)}^{eff}$ is weaker, while the relative strength of the Rashba in-plane field is correspondingly enhanced. This allows the Rashba field to tilt the spinor more efficiently toward the plane, yielding a significantly larger in-plane spin expectation value and, consequently, a substantially stronger field-like SOT in the MoSe$_2$/CrI$_3$ bilayer.

Since no experimental measurements or ab-initio calculations are currently available to suggest a numerical value for the Rashba spin-orbit coupling parameter $\lambda_R$ in these systems, we have carried out our analysis using two different values of $\lambda_R$. This allows us to demonstrate how the choice of Rashba spin-orbit coupling parameter can influence the results. Figures \ref{Fig:2}(c) and \ref{Fig:2}(d) illustrate the calculated SOTs in both n- and p-doped bilayers. For larger $\lambda_R$, both intraband and interband SOTs increase substantially in all cases except n-doped MoSe$_2$, where the sign change of the intraband torque disappears and the stronger interband torque mostly corresponds to the smaller $\lambda_R$. This is mostly related to the distinct geometrical features of the wave function in this case which is the direct consequence of the negative intrinsic spin-orbit coupling, $\lambda_c$, in conduction band, such that if we assumed a positive spin-orbit coupling with the same numerical value, the interband torque would treat as in WSe$_2$-based system and would reduce for smaller Rashba spin-orbit coupling, $\lambda_R$. We should note that the interband term is the result of the modifications of the wave function [see Eq.~\eqref{S_in}] and through that it is strongly linked to symmetry breakings and quantum geometrical characteristics of the wave function. Therefore we can not trace back this effect in the intraband spin-orbit torque in MoSe$_2$/CrI$_3$ bilayer, because as discussed before, this term is mostly related to spin expectation value. The enhancement of the intraband SOT with increasing $\lambda_R$ originates from the direct dependence of the in-plane components of the spin expectation value $\bm{s}^{\chi}$ on $\lambda_R$, in addition to their complex dependence, which can be obtained from Eqs. (\ref{spinexpectation1}) and (\ref{spinexpectation2}). In fact, a larger $\lambda_R$ amplifies the in-plane Rashba field $B_R$, which more effectively cants the spinor toward the plane and thereby increases the in-plane spin polarization that drives the intraband torque. In n-doped MoSe$_2$, Fig. \ref{Fig:12}(b) further shows that increasing $\lambda_R$ broadens the peak structure of the spin expectation value and the resulting spin polarization. This broadening smears out the competition between the two subbands and ultimately suppresses the SOT sign reversal [see Appendix \ref{sec:appendix D}].

In the absence of Rashba SOC, we find that both intraband and intrinsic interband transitions yield zero contribution in the linear response regime and the non-linear spin polarization arising from intraband transitions produces a strong in-plane field-like SOT, nearly an order of magnitude larger than the linear SOT observed in the presence of Rashba SOC [see Appendix \ref{sec:appendix C}]. Finally, we note (not shown) that lowering the temperature below room temperature does not qualitatively alter the chemical potential dependence of the generated SOT components, but merely reduces their magnitudes.

\begin{figure}[t]
\begin{center}
\includegraphics[width=3.7in]{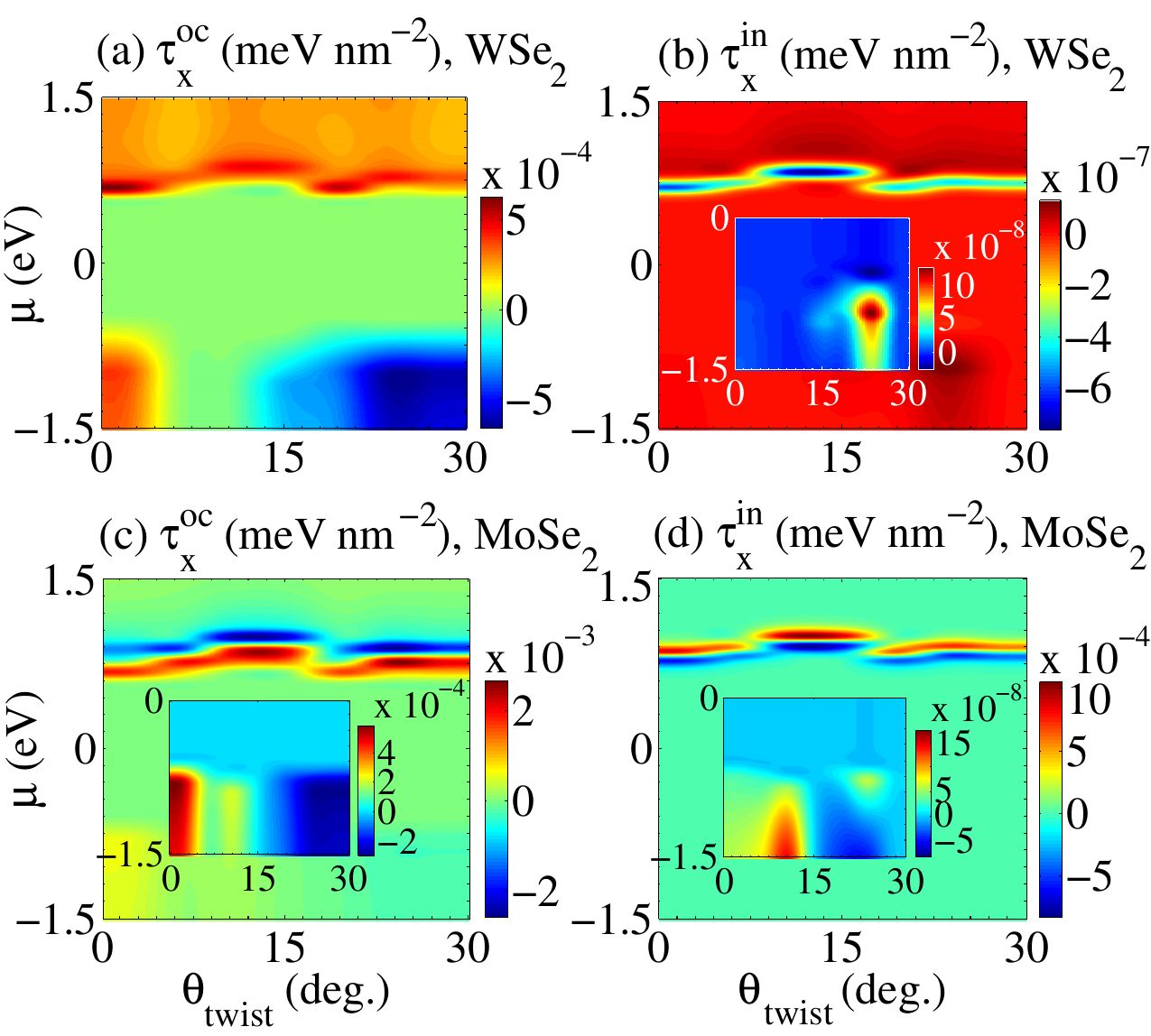}
\end{center}
\caption{\label{Fig:3} (Color online) The colormap of the strength of the spin-orbit torque ${\tau}_x^{oc}$ (left) and ${\tau}_x^{in}$ (right) as functions of the twist angle and the chemical potential for WSe$_2$/CrI$_3$ [(a), (b)] and MoSe$_2$/CrI$_3$ [(c), (d)] bilayers, when $\theta_m =0$ and $\theta_E = \pi/4$. The insets show the zoomed-in view of ${\tau}_x^{oc}$ or ${\tau}_x^{in}$ in the case of p-doped bilayer. The plots are interpolated from the data obtained in Ref. \onlinecite{Zollner23}.}
\end{figure}

\begin{figure}[t]
\begin{center}
\includegraphics[width=3.6in]{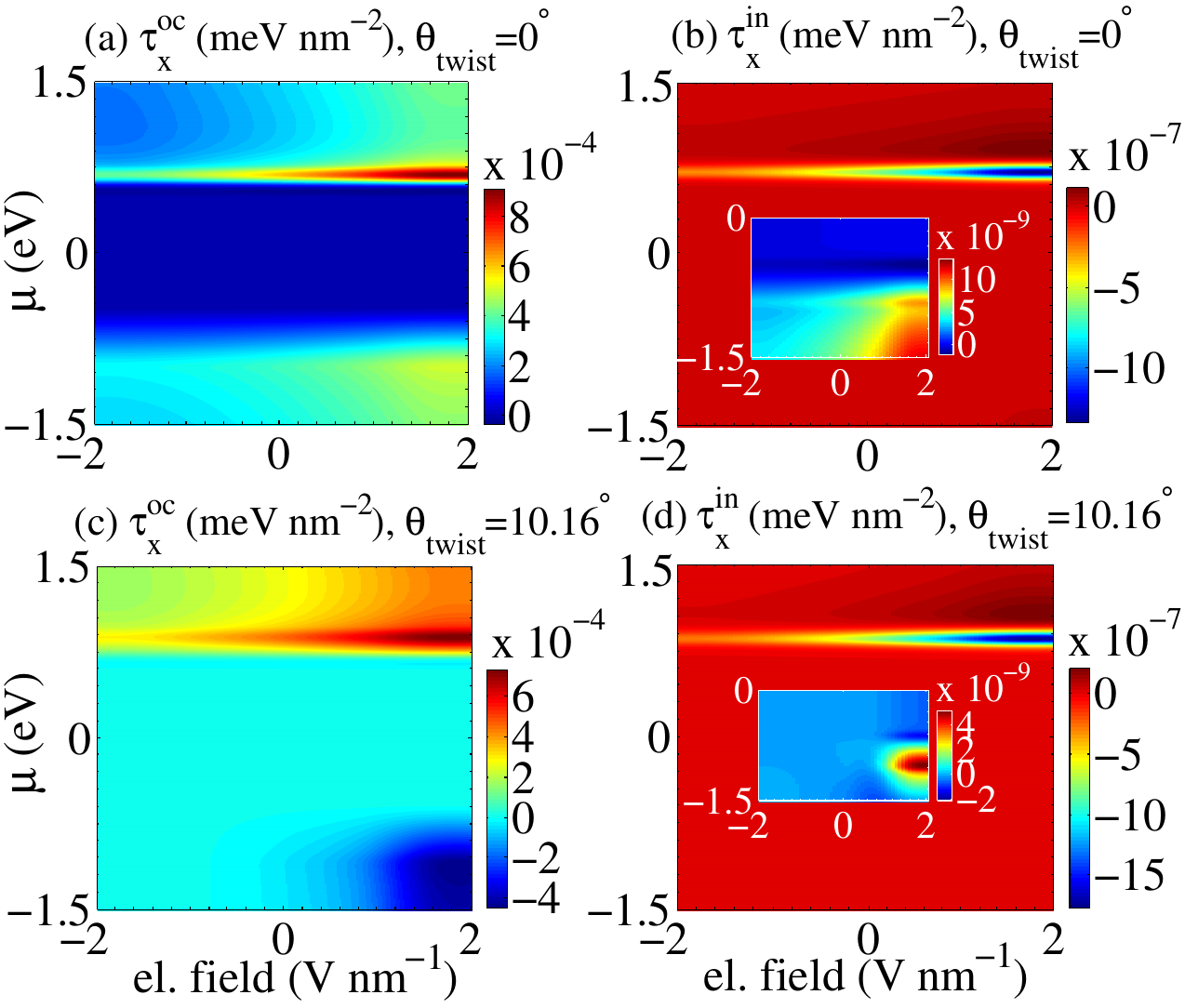}
\end{center}
\caption{\label{Fig:4} (Color online) Dependence of the strength of the spin-orbit torque ${\tau}_x^{oc}$ (left) and ${\tau}_x^{in}$ (right) on the gate electric field and the chemical potential for WSe$_2$/CrI$_3$ bilayer in the absence [(a), (b)] and presence [(c), (d)] of twisting (with $\theta_{twist}=10.16^{\circ}$), when $\theta_m =0$ and $\theta_E = \pi/4$. The insets show the zoomed-in view of the p-doped cases. The plots are interpolated from the data obtained in Ref. \onlinecite{Zollner23}.}
\end{figure}

$Twisting:$In the following, we highlight the extraordinary tunability of the SOT by the effects of the twist-angle. To achieve qualitative understanding of how the SOT is influenced by twisting in a wide range of the twist angle,  in Fig. \ref{Fig:3}, we present the colormap of the strength of the $x$-component of the SOTs as a function of the twist angle $\theta_{twist}$ and the chemical potential $\mu$ for WSe$_2$- and MoSe$_2$-based bilayers, when $\theta_m$ is set to zero, and $\theta_E=\pi/4$. We use the parameters obtained in Ref. \onlinecite{Zollner23} for 7 different twist angles between $0^\circ$ and $30^\circ$, and interpolate the resulting plots using "bicubic" interpolation method. Importantly, twisting the TMDC layer relative to the ferromagnetic CrI$_3$ layer produces pronounced modifications in the SOTs. This arises predominantly from the twist-dependent evolution of the proximity exchange parameters $B_c$ and $B_v$ reported in Ref. \onlinecite{Zollner23}, which directly modulate both intra and interband SOTs-notably the zero-SOT gap and the sharp peak in n-doped systems- making them to track these changes accordingly since $\bm{\tau}\propto B_{c(v)}$ [see Eq. (\ref{torque})]. Moreover, the twist-dependent evolution of the gap parameter $\Delta$ plays a key role in determining the $\chi$-band edge energies and the resulting zero-SOT gap [see Fig. \ref{Fig:9} in Appendix \ref{sec:appendix B}]. Therefore, a large discrepancy between the n-type and p-type doping can be apparently seen for both TMDC/CrI$_3$ bilayers. Tuning the twist angle between $0^{\circ}$ and $30^\circ$ induces substantial changes in the magnitude of the intraband spin-orbit torque $\tau_x^{oc}$, accompanied by a sign reversal in p-doped WSe$_2$- and MoSe$_2$-based bilayers, because of the sign change of $B_v$, and two distinct sign changes in the n-doped MoSe$_2$/CrI$_3$ bilayer, depending on the chemical potential [see Figs. \ref{Fig:3}(a) and \ref{Fig:3}(c)]. As explained before, this discrepancy between the two TMDCs mainly arises from different sign of the spin-orbit coupling parameters in their conduction bands. Most notably, as we also saw before, the $\tau_x^{oc}$ strength in MoSe$_2$-based bilayer is approximately one order of magnitude larger than in WSe$_2$ bilayer. The intrinsic spin-orbit torque $\tau_x^{in}$ is likewise highly sensitive to twisting of the ferromagnetic layer. As shown in Figs. \ref{Fig:3}(b) and \ref{Fig:3}(d), the torque magnitude can change abruptly, with variations exceeding one order of magnitude at certain energies and twist angles, depending on the material and doping type. Remarkably, twisting also induces sign reversals of $\tau_x^{in}$ in both MoSe$_2$ and WSe$_2$ bilayers for n- and p-doped regimes. Most importantly, the obtained spin-orbit torque ratio $\tau_x^{oc}/\tau_x^{in}$ is on the order of $10^3$ for the WSe$_2$/CrI$_3$ bilayer and reaches approximately $10$ for the MoSe$_2$-based structure.

\begin{figure}[t]
\begin{center}
\includegraphics[width=3.6in]{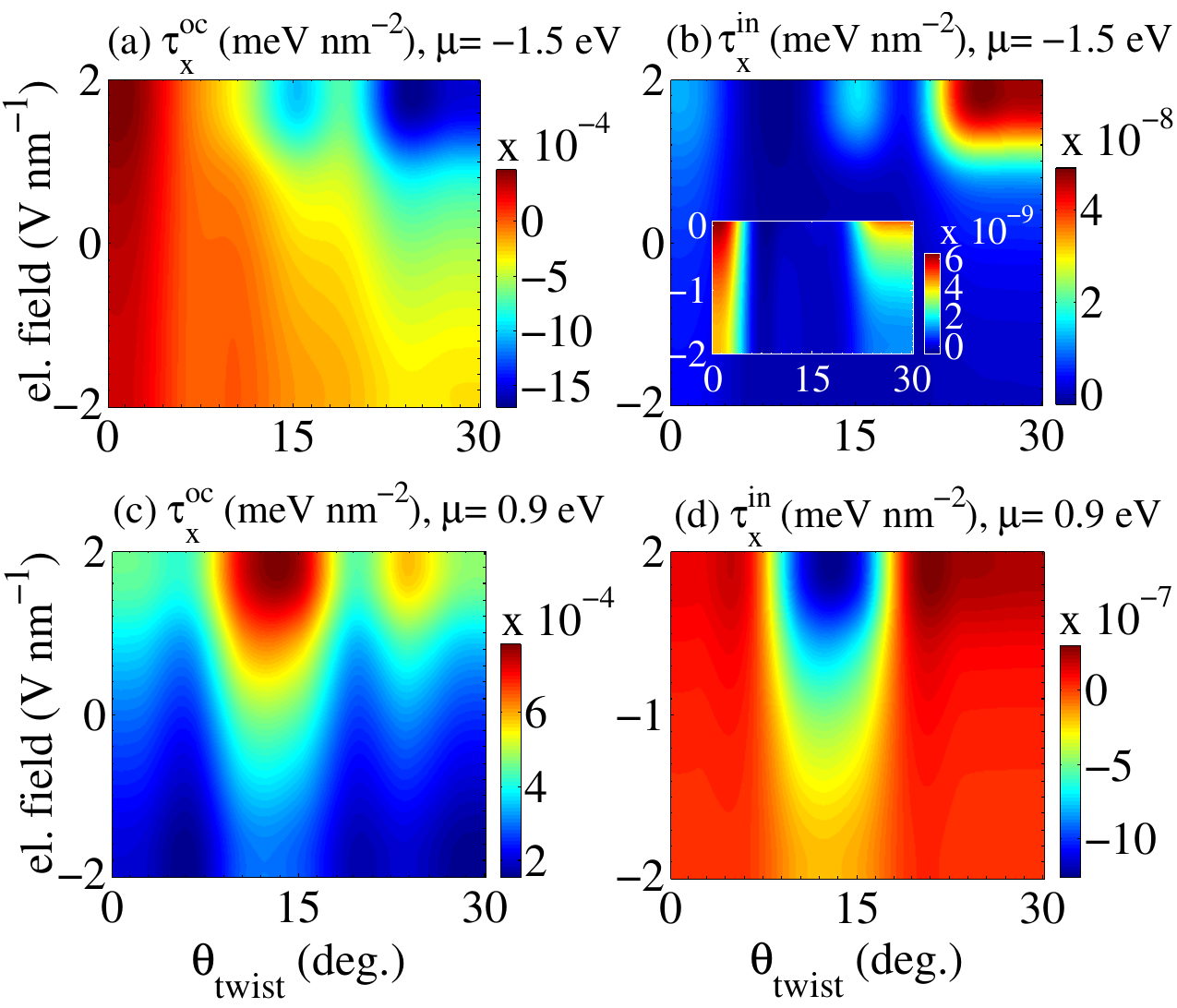}
\end{center}
\caption{\label{Fig:5} The colormap of the strength of the spin-orbit torque ${\tau}_x^{oc}$ (left) and ${\tau}_x^{in}$ (right) as functions of the twist angle and the gate electric field for two values of the chemical potential $\mu = -1.5$ eV [(a), (b)] and $0.9$ eV [(c), (d)] in WSe$_2$/CrI$_3$ bilayer, when $\theta_m = 0$ and $\theta_E=\pi/4$. Inset of (b) shows the zoomed-in view of the negative gate electric field case. The plots are interpolated from the data obtained in Ref. \onlinecite{Zollner23}.}
\end{figure}
The colormap of the strength of the spin-orbit torque ${\tau}_x^{oc}$ (left) and ${\tau}_x^{in}$ (right) as functions of the twist angle and the gate
electric field for two values of the chemical potential $\mu = -1.5$ eV [(a), (b)] and $0.9$ eV [(c), (d)] in WSe$_2$/CrI$_3$ bilayer

$Gating:$There is also a possibility to tune both the strength and the sign of the SOTs by applying a transverse electric field of a few V$/$nm across the twisted TMDC/CrI$_3$ bilayer. Figure \ref{Fig:4} presents the results of our calculation for WSe$_2$-based bilayer in two cases: with no twist and a $10.16^{\circ}$ twist. The gate electric field directly affects the proximity exchange parameters $B_c$ and $B_v$. While for all twist angles a gate scan from $-2$ V$/$nm to $2$ V$/$nm only enhances the absolute value of the exchange parameters, there is a sign reversal in $B_v$ for the special case of $\theta_{twist}=10.16^{\circ}$ [see Ref. \onlinecite{Zollner23}]. The intraband and intrinsic interband SOTs also follow the changes in proximity parameters. A gate sweep from $-2$ V$/$nm to $2$ V$/$nm increase the torque magnitudes, ${\tau}_x^{oc}$ and ${\tau}_x^{in}$, in n-doped systems for both $0^{\circ}$ and $10.16^{\circ}$ twist, as well as in p-doped systems for $0^{\circ}$ twist. Remarkably, in the p-doped system with a $10.16^{\circ}$ twist, we find that gating can also induce a sign reversal of the torques ${\tau}_x^{oc}$ and ${\tau}_x^{in}$.

Finally, the striking tunability of the spin-orbit torques ${\tau}_x^{oc}$ and ${\tau}_x^{in}$ through the combined effects of twist angle and gate electric field can be of particular interest. This is illustrated in Fig. \ref{Fig:5} for WSe$_2$-based bilayer with both n- and p-type doping. Remarkably, significant amplification of the SOTs strengths, as well as their sign reversal in p-doped system, can be achieved across a broad range of twist angles and gate electric fields. Interestingly, the sign change of the intrinsic SOT is also possible in the case of n-doped bilayer, with the torque strength remaining comparable to that of the p-doped case. It is important to note that the intraband and interband contributions to the SOT exhibit opposite signs.

\section{conclusion}\label{conclusion}

In summary, we have studied the current-induced spin polarization and the resulting spin-orbit torque (SOT) in a transition metal dichalcogenide (TMDC) and chromium iodide (CrI$_3$) bilayer, with WSe$_2$ or MoSe$_2$ as TMDC, beyond the linear response theory. Employing the single-band steady state Boltzmann equation, we have found that the spin-polarization originating from the electron occupation changes within intraband is odd upon magnetization reversal and generates a strong field-like SOT, $\bm{\tau}^{oc}$, on the magnetization of the ferromagnetic CrI$_3$ layer, while that of the intrinsic interband transitions contribute to a weak damping-like torque, $\bm{\tau}^{in}$, in TMDC/CrI$_3$ bilayers. Both the field-like and damping-like terms exhibit strongly asymmetric behavior with respect to the chemical potential with higher strength in the case of n-type doping and sign changes for n-doped MoSe$_2$/CrI$_3$ bilayers.

We have evaluated the effects of the type of the TMDC, twisting the TMDC layer with respect to the CrI$_3$, and the gate electric field on the generated SOT. Interestingly, the strength of the damping-like (field-like) SOT in MoSe$_2$-based bilayer can be three orders (one order) of magnitude larger than that of the WSe$_2$-based bilayer, leading to an approximately the same order as the field-like torque. Moreover, the field-like and damping-like SOTs are enhanced for larger Rashba spin-orbit couplings in WSe$_2$- and p-type doped MoSe$_2$-based bilayers, whereas in n-doped MoSe$_2$, the strength of the damping-like SOT decreases and the sign change of the field-like torque is suppressed.

Twisting the TMDC layer with respect to the ferromagnetic CrI$_3$ layer leads to significant changes in applied current-induced SOTs. Importantly, a large discrepancy between the n-type and p-type doping can be apparently seen in WSe$_2$ and MoSe$_2$-based bilayers. Not only do the magnitudes of the SOTs can change significantly, but also the sign of the damping-like SOT as well as the field-like SOT (except that of the n-doped WSe$_2$) can be reversed by tuning the twist angle between $0^{\circ}$ and $30^{\circ}$ in both TMDCs.

Furthermore, tuning the applied gate electric field from $-2$ V/nm to $2$ V/nm can lead to the sign change of the SOTs in addition to the amplification of their strengths. Most importantly, we have depicted the striking tunability of the strength of the generated SOTs as well as their sign change in a wide window of the twist angle and the applied gate electric field which opens a new path for externally controllable spintronic devices.

Since several works in the field of SOTs using TMDC-based devices are available, a proper comparison with those results seems to be in order. Experimental studies show that various factors- including the source of TMDCs (e.g., mechanical exfoliation or chemical vapor deposition), the choice of ferromagnetic materials (FM), deposition techniques (such as sputtering or electron-beam evaporation), and measurement techniques like second-harmonic Hall (SHH)~\cite{Garello2013,Hayashi2014,Avci2014,Ghosh2017} or spin-torque ferromagnetic resonance (ST-FMR)~\cite{Liu2011,Fang2011,Berger2018}- affect the measured SOTs.

Shao $\textit{et al.}$ reported an absence of in-plane damping-like torque and a temperature-independent out of plane field-like torque $\tau_{FL}\sim 10^{-7}$ ($10^{-8}$) meV/nm$^{2}$ in monolayer WSe$_2$ (MoS$_2$) coupled with CoFeB, quantified using non-resonant SHH measurements~\cite{shao2016strong}. Interestingly, in a concurrent study, Zhang $\textit{et al.}$ obtained different results using a high-frequency technique, ST-FMR, on monolayer MoS$_2$/Permalloy (Ni$_{80}$Fe$_{20}$–Py)~\cite{zhang2016research} and detected both in-plane damping-like and out-of-plane field-like torques, with a torque ratio of $\tau_{FL}/\tau_{DL} = 0.19\pm 0.01$, indicating that the damping-like torque dominates over the field-like torque. Crucially, this result repeated using different deposition techniques of the FM layer deposition methods, indicating that the observed torque is not affected by the Py deposition technique. While different measurement techniques might explain the discrepancy in the observed torques, it has been demonstrated that SOT values quantified by ST-FMR and SHH techniques are typically in good agreement within experimental accuracy across several systems~\cite{macneill2017control,stiehl2019current,Shi2019}. The conflicting findings in MoS$_2$/FM bilayers highlight that not only the spin-orbit material but also the type of ferromagnetic layer plays a significant role in the resulting torque characteristics. This is theoretically supported by recent calculations on heterostructures such as MoSe$_2$/Co, WSe$_2$/Co, and TaSe$_2$/Co~\cite{dolui2020spin}. Moreover, Lv $\textit{et al.}$ observed both damping-like and field-like torques in ST-FMR measurements on monolayer WS$_2$/Py heterostructures~\cite{lv2018electric}, using CVD-grown WS$_2$ and electron-beam evaporated Py, attributed to the interfacial Rashba-Edelstein effect. Notably, they observed a gate-dependent SOT ratio ranging from $\tau_{FL}/\tau_{DL} = 0.05$ to $0.22$ within a range of $-60$ V to $+60$ V, an effect absent in their Pt/Py reference samples.

To the best of our knowledge, this is the first report to explore SOT in TMDCs interfaced with two-dimensional (2D) FM CrI$_3$, revealing both field-like and damping-like torque components. Remarkably, the strength of both field-like and damping-like torques are at least three orders of magnitudes larger than previously reported results. The obtained spin torques ratio $\tau_{FL}/\tau_{DL}$, reaching approximately 10$^3$ for WSe$_2$ and 10 for MoSe$_2$, exceed prior studies by two to four orders of magnitude. It is worth noting that the predicted magnitudes of the spin torques and the resulting $\tau_{FL}/\tau_{DL}$ ratio in MoSe$_2$/CrI$_3$ bilayer are comparable to those obtained in TI/2D FM bilayer with a value of approximately $\tau_{FL}/\tau_{DL}\sim 10^{-3}/10^{-4}=10$ ~\cite{Ghosh2017a}, whereas the corresponding ratio in WSe$_2$-based bilayer is significantly higher than the value reported in TI/2D FM system. These findings highlight CrI$_3$ as a highly promising 2D FM for externally controllable spintronic devices.
\appendix

\section {\label{sec:appendix A} The applied current-generated non-equilibrium intraband spin polarization}

\begin{figure}[t]
\begin{center}
\includegraphics[width=3.1in]{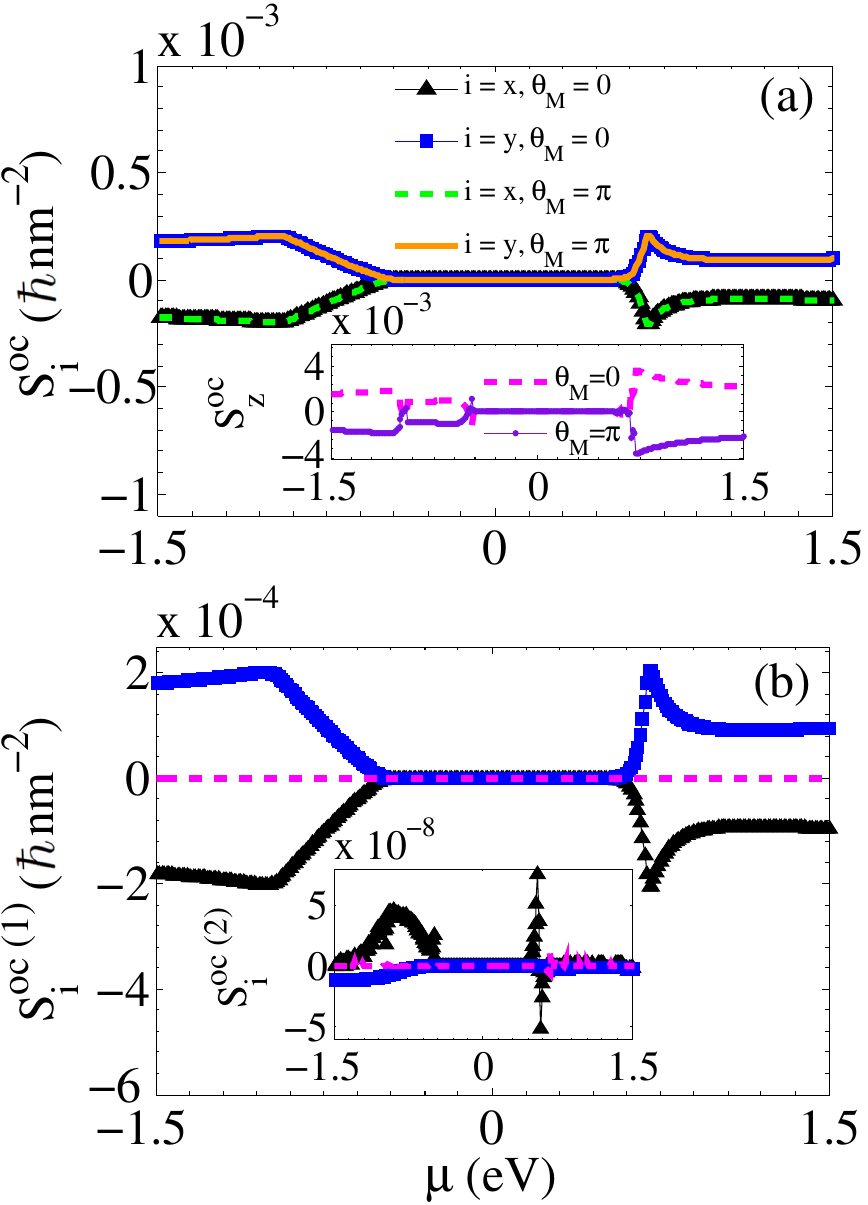}
\end{center}
\caption{\label{Fig:6} (Color online) (a) Chemical potential dependence of the non-equilibrium intraband spin polarization $\bm{S}^{oc}$ components in WSe$_2$/CrI$_3$ bilayer for two magnetization directions $\theta_m=0$ and $\pi$, with $\lambda_R = 1$ meV, $\theta_{twist}=0^\circ$ and $\theta_E=\pi/4$. (b) Contributions of the linear $\bm{S}^{oc(1)}$ and non-linear $\bm{S}^{oc(2)}$ terms to the spin polarization components, for $\theta_m=0$. The labels in panel (b) are the same as  those introduced in panel (a).}
\end{figure}
\begin{figure}[t]
\begin{center}
\includegraphics[width=3.4in]{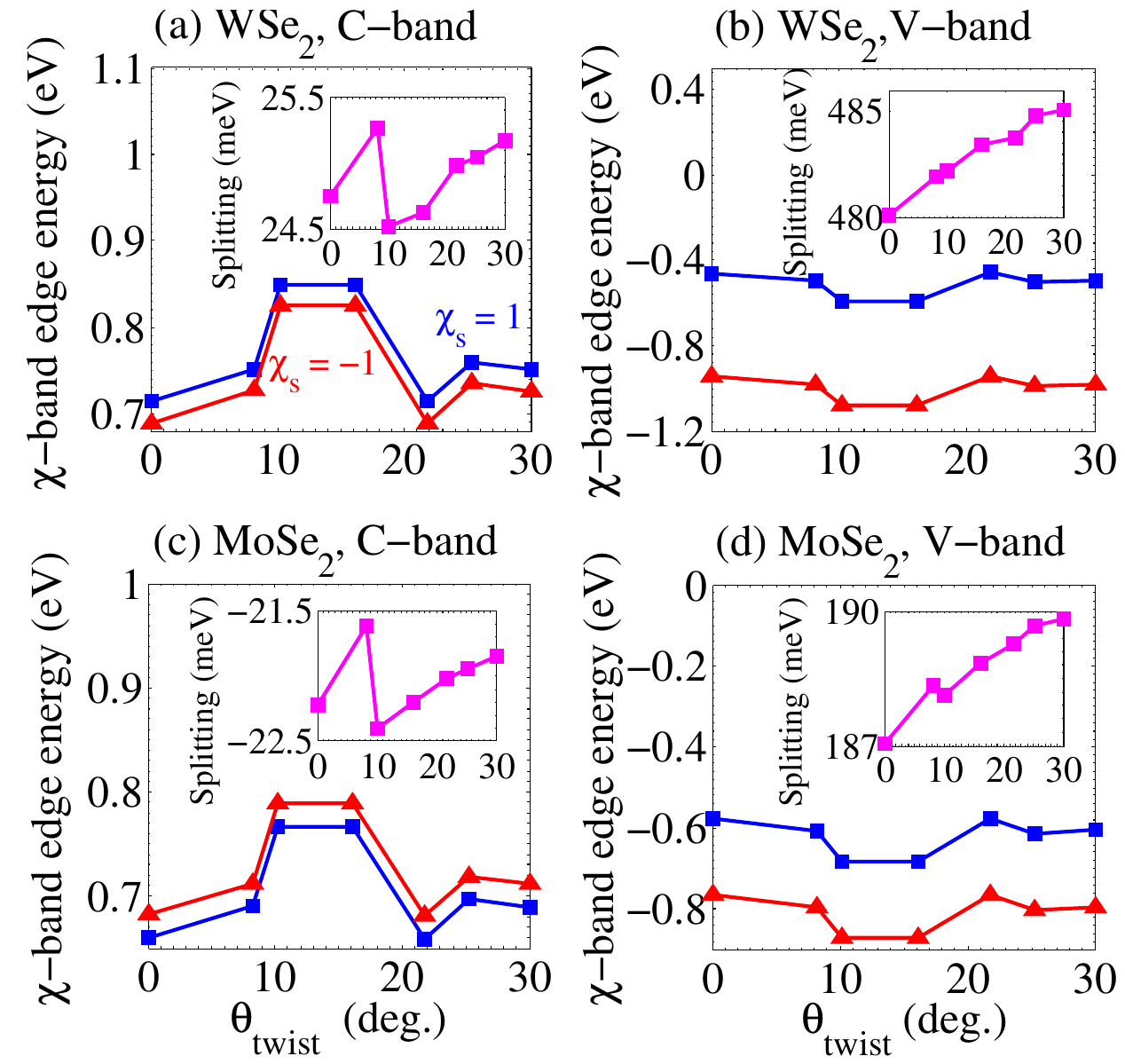}
\end{center}
\caption{\label{Fig:9} (Color online) Top (Bottom) panel shows the twist-angle dependence of $\chi_s = \pm 1$ subband edge energies in the conduction (C) and valence (V) bands of the K valley, with $\chi_{\tau}=1$, for WSe$_2$/CrI$_3$ and MoSe$_2$/CrI$_3$ bilayers. The insets display the corresponding splitting energies of the $\chi_s = 1$ and $\chi_s = -1$ subbands in terms of the twist-angle. Subbands with $\chi_s=1$ and $-1$ are indicated by blue lines with square markers and red lines with triangle markers, respectively, following the labeling convention introduced in panel (a). The magnetization direction is fixed at $\theta_m=0$, and Rashba SOC parameter is set to $\lambda_R=1$ meV.}
\end{figure}
Figure \ref{Fig:6}(a) presents the behavior of the current-induced spin polarization $\bm{S}^{oc}$ components stemming from the intraband transitions in terms of the chemical potential $\mu$ in WSe$_2$/CrI$_3$ heterostructure, when $\theta_{twist}=0$, $\lambda_R = 1$ meV, and $\theta_E=\pi/4$. It is found that the in-plane components of the intraband spin polarization, $S_x^{oc}$ and $S_y^{oc}$, are even with respect to the magnetization direction, whereas the out of plane component $S_z^{oc}$ is odd function of $\theta_{m}$ and one order of magnitude larger than those of the in-plane components. The applied current-generated non-equilibrium spin polarization components exhibit strongly asymmetric behavior with respect to the chemical potential, without any change of sign, and exert an in-plane spin-orbit torque $\bm{\tau}^{oc}=2 J \hbar^{-1} m_z(-S_y^{oc},S_x^{oc},0)$, with respect to the WSe$_2$/CrI$_3$ bilayer, on the magnetization of the FM layer.

The contribution of the linear $\bm{S}^{oc(1)}=\sum_{\chi}\int\frac{d^2\bm{k}}{(2\pi)^2}\bm{s}^{\chi}(\bm{k})f^{(1)}(\varepsilon_{\bm{k}}^{\chi})$ and non-linear $\bm{S}^{oc(2)}=\sum_{\chi}\int\frac{d^2\bm{k}}{(2\pi)^2}\bm{s}^{\chi}(\bm{k}) f^{(2)}(\varepsilon_{\bm{k}}^{\chi})$ terms to the intraband spin polarization are depicted in Fig. \ref{Fig:6}(b). It is seen that the non-linear term is approximately four orders of magnitude smaller than the linear term. Consequently, the spin polarization $\bm{S}^{oc}$, and thus the resulting spin-orbit torque, is dominated by the linear response regime. In contrast, due to the much weaker magnitude of the intrinsic interband spin polarization $\bm{S}^{in}$ and the resulting spin-orbit torque in comparison with those of the intraband transitions, the second-order response is negligible in this case and can be safely disregarded~\cite{zhou2022nonlinear}.

\section {\label{sec:appendix B} Twist-angle dependence of electronic subband-edge energies in TMDC/CrI$_3$ bilayers}
\begin{figure}[]
\begin{center}
\includegraphics[width=3.4in]{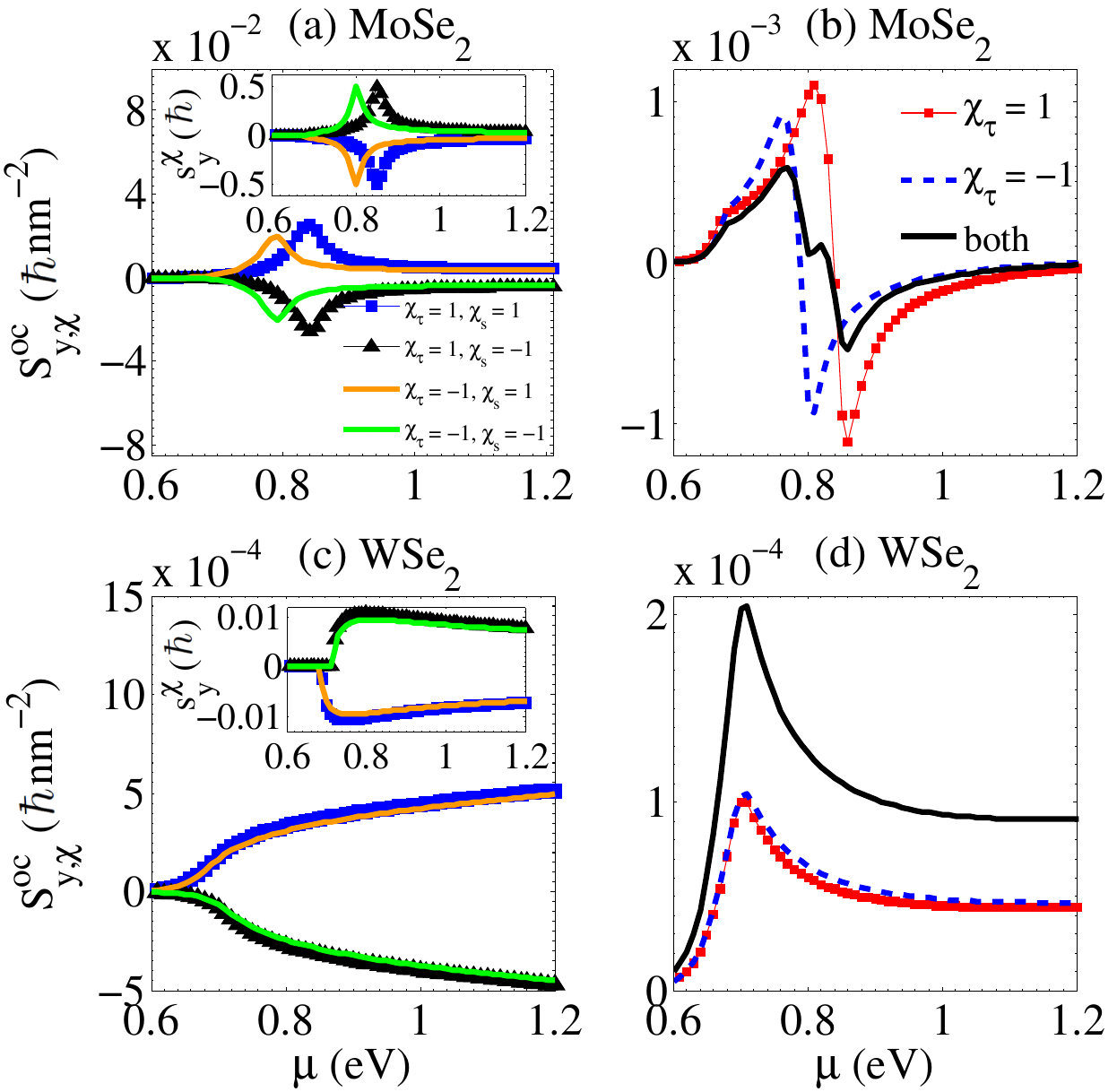}
\end{center}
\caption{\label{Fig:11} Chemical potential dependence of the intraband spin polarization ${S}_{y, \chi}^{oc}$ for each $\chi_s=\pm 1$ subband (left) and for their combined contribution (right) within the K and K' valleys (with $\chi_{\tau}=\pm 1$) of the MoSe$_2$/CrI$_3$ [(a), (b)] and WSe$_2$/CrI$_3$ [(c), (d)] bilayers, when $\theta_m=0$, $\lambda_R = 1$ meV, $\theta_{twist}=0^\circ$ and $\theta_E=\pi/4$. Insets of (a) and (c) present the behavior of the corresponding spin expectation value $s_y^{\chi}$ in the $\chi=(\chi_{\tau},\chi_s)$ subband at $\bm{k}=0$. The labels in panels (c) and (d) are the same as those introduced in panels (a) and (b), respectively.}
\end{figure}
Figure \ref{Fig:9} depict the twist-angle dependence of the $\chi_s = \pm 1$ subband edge energies in the conduction and valence bands of the K valley, together with the associated splitting energies for bilayers based on WSe$_2$ and MoSe$_2$. Our analysis reveals that the evolution of the subband splitting energies with twist angle exhibits distinct material- and band-dependent trends. In WSe$_2$ bilayers, the splitting energies of the $\chi_s = 1$ and $-1$ subbands in both the conduction and valence bands increase almost linearly as the twist angle grows, except around the twist angle $\theta_{twist} = 10.16^{\circ}$. In contrast, MoSe$_2$ bilayers display a more asymmetric response: while the splitting energies in the valence band increase almost linearly with twist angle, those in the conduction band decrease with increasing twist angle (arising from the different signs of $\lambda_c$ and $B_c$), except around $\theta_{twist} = 10.16^{\circ}$. An opposite trend is observed in the behavior of the splitting energies of the $\chi_s = 1$ and $-1$ subbands in the K' valley. Overall, the observed twist-angle dependence of the subband splitting energies in both WSe$_2$- and MoSe$_2$-based bilayers is consistent with the behavior of the proximity exchange coupling of the conduction and valence bands reported in Ref. \onlinecite{Zollner23}. Anomalous behavior at a twist angle of $10.16^{\circ}$ is linked to the amplification of the proximity exchange field within the conduction band.

\section {\label{sec:appendix D} Contribution of the carriers from different subbands to the intraband spin polarization}

\begin{figure}[]
\begin{center}
\includegraphics[width=3.5in]{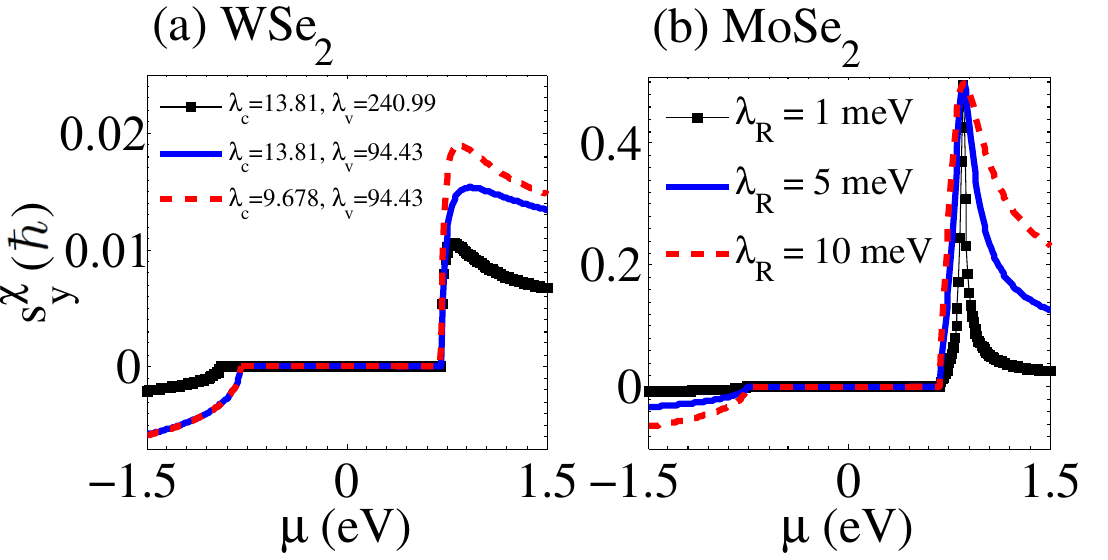}
\end{center}
\caption{\label{Fig:12} (Color online) Chemical potential dependence of the spin expectation value $s_y^{\chi}$ for WSe$_2$/CrI$_3$ bilayer with different values of the intrinsic spin-orbit coupling parameters $\lambda_c$ and $\lambda_v$ (in units of meV) (a) and MoSe$_2$/CrI$_3$ bilayer with three values of the Rashba spin-orbit coupling parameter $\lambda_R$ (b), when $\chi=(\chi_{\tau}=1,\chi_s=-1)$, $\bm{k}=0$, $\theta_m=0$, $\theta_{twist}=0^\circ$ and $\theta_E=\pi/4$. We should note that the intrinsic spin-orbit coupling parameters in the conduction and valence bands of WSe$_2$/CrI$_3$ (MoSe$_2$/CrI$_3$) bilayers are $\lambda_c= 13.81$ ($-9.678$) meV and $\lambda_v= 240.99$ ($94.43$) meV.}
\end{figure}
To elucidate the chemical potential dependence of the intraband spin-orbit torque $\bm{\tau}^{oc}$ in TMDC/CrI$_3$ bilayers, we analyze the contribution of carriers from distinct subbands to the spin polarization $\bm{S}^{oc}$. Assuming the local magnetization $\bm{m}$ is aligned along the z-axis, the intraband spin-orbit torque is given by $\bm{\tau}^{oc}=2 J {\hbar}^{-1} (-S_y^{oc}, S_x^{oc},0) $. So the x-component (equal to $ {\tau}_y^{oc}$  for $\theta_E=\pi/4$) scales linearly with $S_y^{oc}$. The top and bottom panels of Fig. \ref{Fig:11} depict the subband-dependent contributions to $S_y^{oc}$ from the $\chi_s = \pm 1$ subbands in the K and K' valleys (with $\chi_{\tau}= \pm 1$) for n-doped MoSe$_2$- and WSe$_2$-based bilayers, respectively, when $\lambda_R = 1$ meV, $\theta_{twist}=0^\circ$ and $\theta_E=\pi/4$. We demonstrate that the spin polarization of the $\chi_s = \pm 1$ subbands within each valley are opposite in sign [see Figs. \ref{Fig:11}(a) and \ref{Fig:11}(c)]. The absence of quasiparticle states within the band gap of the TMDC/CrI$_3$ bilayer generates a gap for energies below the $\chi$-band edge with $\chi=(\chi_{\tau},\chi_s $) [see Fig. \ref{Fig:00}], and a peak appears for the n-doped case. This peak is predominantly governed by the spin expectation value $s_y^{\chi}$ of carriers in the corresponding subband, as shown in the insets of Figs. \ref{Fig:11}(a) and \ref{Fig:11}(c).

Importantly, the width and amplitude of the peaks are highly sensitive to the signs and magnitudes of intrinsic and Rashba spin-orbit coupling parameters. According to Table \ref{table1}, the intrinsic spin-orbit coupling parameters in the conduction and valence bands of MoSe$_2$/CrI$_3$ (WSe$_2$/CrI$_3$) bilayers are $\lambda_c=-9.678$ ($13.81$) meV and $\lambda_v=94.43$ ($240.99$) meV. In MoSe$_2$/CrI$_3$ bilayer, the negative $\lambda_c$ yields a sharp, high-amplitude peak in the spin polarization of each $\chi$-subband. The resulting sum of opposing spin polarizations from the $\chi_s = \pm 1$ subbands makes a sign change in the spin polarization of the K and K' valleys, thereby reversing the net spin polarization of the n-doped system [see Figs. \ref{Fig:11}(a) and \ref{Fig:11}(b)]. In contrast, the positive $\lambda_c$ in WSe$_2$-based bilayer produces a broad, low-amplitude peak in $S_{y, \chi}^{oc}$, suppressing the sign reversal in each valley and the resulting net spin polarization [see Figs. \ref{Fig:11}(c) and \ref{Fig:11}(d)]. Moreover, as is shown in Fig. \ref{Fig:12}(a), the larger values of $\lambda_c$ and specially $\lambda_v$ in WSe$_2$ relative to MoSe$_2$, further reduces the spin polarization, thereby enhancing the intraband SOT in MoSe$_2$-based bilayer. Finally, the chemical potential at which the peak in intraband spin polarization and associated SOT occurs is determined by the contributions of the both valleys. In addition, Fig. \ref{Fig:12}(b) demonstrates that increasing the Rashba spin-orbit coupling $\lambda_R$ not only amplifies the spin expectation value and its associated spin polarization but also broadens the peak. This broadening can obscure the subband opposition, ultimately suppressing the sign change in the spin polarization and the resulting intraband SOT for n-doped MoSe$_2$.

\section {\label{sec:appendix C} Spin-orbit torque in the case of zero Rashba spin-orbit coupling}
\begin{figure}[]
\begin{center}
\includegraphics[width=3.6in]{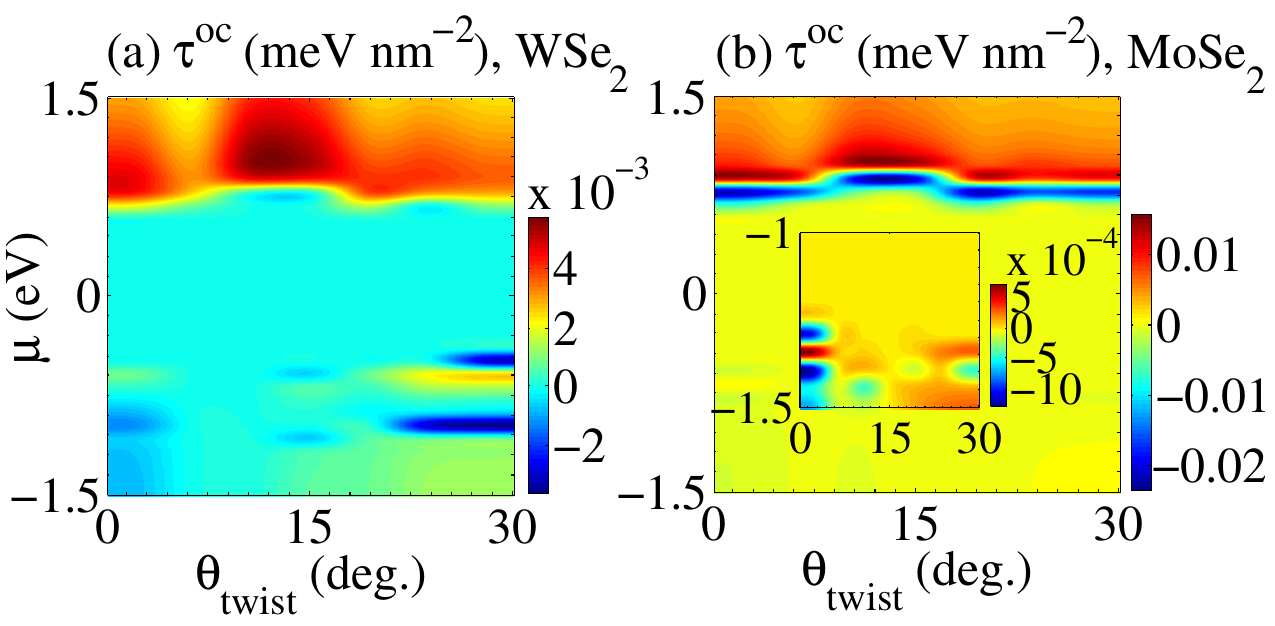}
\end{center}
\caption{\label{Fig:10} (Color online) The colormap of the strength of the spin-orbit torque ${\tau}^{oc}$ as a function of the twist angle and the chemical potential for (a) WSe$_2$/CrI$_3$ and (b) MoSe$_2$/CrI$_3$ bilayers, when $\lambda_R=0$, $\theta_m =0$ and $\theta_E = \pi/4$. Inset of (b) shows the zoomed-in view of ${\tau}^{oc}$ in the case of p-doped MoSe$_2$-based bilayer. The plots are interpolated from the data obtained in Ref. \onlinecite{Zollner23}.}
\end{figure}
We evaluate the results in the absence of Rashba SOC parameter. Importantly, we find that the contribution of the intraband and intrinsic interband transitions is zero in linear response regime, and the non-linear spin polarization associated with intraband transitions generates a strong in-plane field-like spin-orbit torque $\bm{\tau}^{oc}=2 J \hbar^{-1}(m_z S_x^{oc}-m_x S_z^{oc})\ \hat{y}$, on the ferromagnetic CrI$_3$ layer. Remarkably, the non-linear field-like torque ${\tau}^{oc}$ is one order of magnitude larger than that of the linear SOT in the presence of Rashba SOC. As depicted in Fig. \ref{Fig:10}, the SOT has asymmetric dependence on the chemical potential with larger value in the case of n-type doping and changes sign for MoSe$_2$- and p-type doped WSe$_2$-based bilayers, in contrast to the case with finite Rashba SOC. Twisting the TMDC layer with respect to the ferromagnetic CrI$_3$ layer results in an amplification or attenuation of the strength of the SOT depending on the chemical potential as well as a sign change in n-type doped WSe$_2$-based bilayer. Moreover, the strength of the spin-orbit torque in MoSe$_2$-based bilayer with n-type doping is found to be one order of magnitude larger than that of the WSe$_2$/CrI$_3$ bilayer. We additionally find that the SOT is insensitive to the direction of the applied in-plane electric field, in contrast to the case with finite Rashba.

\end{document}